\pdfoutput=1

\documentclass[a4paper,10pt]{article}
\usepackage{jheppub,amsmath,amsfonts,amssymb,xcolor,graphicx,hyperref}
\usepackage[utf8x]{inputenc}
\usepackage[section]{placeins}
\usepackage{cancel}

\usepackage{subfigure,color,graphicx,hyperref,dsfont,slashed,multirow}
\usepackage{fontenc}
\usepackage{pbox}
\usepackage{cancel}
\usepackage{setspace}
\usepackage{float}
\usepackage{mathtools,nccmath}%
\usepackage{ etoolbox, xparse} 
\usepackage{array}
\usepackage[normalem]{ulem}
\usepackage{amstext}
\usepackage{booktabs}

\hypersetup{colorlinks=true,linkcolor=blue,citecolor=magenta,filecolor=magenta,
	urlcolor=cyan}
\newcommand{\mg}{\textsc{MG5\_aMC}}
\newcommand{\ma}{\textsc{MadAnalysis}~5}
\newcommand{\mo}{\textsc{MicrOMEGAs}}

\newcommand{\spheno}{\textsc{SPheno}}

\newcommand{\sa}{\textsc{SARAH}}
\newcommand{\be}{\begin{equation}}
	\newcommand{\ee}{\end{equation}}
\def\bsp#1\esp{\begin{split}#1\end{split}}
\def\bpm{\begin{pmatrix}}
	\def\epm{\end{pmatrix}}
\newcommand{\bea}{\begin{eqnarray}}
	\newcommand{\eea}{\end{eqnarray}}

\title{Electron and muon magnetic moments and implications for dark matter and model characterisation in non-universal $U(1)^\prime$ supersymmetric models\footnote{Dedicated to the memory of our colleague Levent Solmaz, who  made many contributions to $U(1)^\prime$ models.}}

\author[a]{Mariana~Frank}
\author[b,c]{\!\!, Ya\c{s}ar Hi\c{c}y\i lmaz}
\author[d]{\!\!, Subhadeep Mondal}
\author[a]{\!\!, \"{O}zer \"{O}zdal}
\author[e]{\! and  Cem Salih \"{U}n}

\emailAdd{mariana.frank@concordia.ca}
\emailAdd{Y.Hicyilmaz@soton.ac.uk}
\emailAdd{Subhadeep.Mondal@bennett.edu.in}
\emailAdd{ozer.ozdal@soton.ac.uk}
\emailAdd{cemsalihun@uludag.edu.tr}

\affiliation[a]{Department of Physics, Concordia University,
	7141 Sherbrooke St. West, Montreal, Quebec, Canada H4B 1R6}
\affiliation[b]{School of Physics $\&$ Astronomy, University of Southampton, Highfield, Southampton SO17 1BJ,UK}
\affiliation[c]{Department of Physics, Bal\i kesir University, TR10145, Bal\i kesir, Turkey}
\affiliation[d]{Bennett University,  Plot Nos 8-11, TechZone II, Greater Noida 201310, Uttar Pradesh, India}
\affiliation[e]{Department of Physics, Bursa Uluda\~{g} University, TR16059, Bursa, Turkey}

\abstract{We attribute deviations of the muon and electron magnetic moments from the theoretical predictions to the presence of an additional  $U(1)^\prime$ supersymmetric model. We interpret the discrepancies between the muon and electron anomalous magnetic moments to be due to the presence of non-universal  $U(1)^\prime$ charges. In a minimally extended model, we show that requiring both deviations to be satisfied imposes constraints on the spectrum of the model, in particular on dark matter candidates and slepton masses and ordering. Choosing three benchmarks with distinct dark matter features, we study implications of the model at colliders, concentrating on variables that can  distinguish our non-universal scenario from other $U(1)^\prime$ implementations.
}

\begin{document}	

	\maketitle
	\flushbottom

\section{Introduction}
\label{sec:intro}

The first preliminary results on the muon anomalous magnetic moment (known as muon $g-2$) from the Muon $g-2$ experiment at Fermilab, were recently revealed as $a_{\mu} = 116592040(54)\times10^{-11}$ \cite{Abi:2021gix,Albahri:2021kmg,Albahri:2021ixb}, which, after combining its results with those from the experiments at the Brookhaven National Laboratory (BNL) \cite{Bennett:2006fi}, led to a new world average \cite{Abi:2021gix}:

\begin{equation}
 a_{\mu}^{{\rm WA}} = 116592061(41)\times 10^{-11}~,
 \end{equation} 
where $a_{\mu} \equiv (g-2)/2$ is the anomalous magnetic moment of the muon, while $g$ denotes the gyromagnetic ratio, which is equal to $2$ at tree-level. 

Comparing this result with the theoretical prediction of the Standard Model (SM) ($a_{\mu}^{{\rm SM}} = 116591810(43)\times10^{-11} $ \cite{Aoyama:2020ynm}) the current experimental results on muon $g-2$ now lead to a deviation of $4.2\sigma$ from the SM prediction, expressed as \cite{Abi:2021gix}

\begin{equation}
\Delta a_{\mu} = a_{\mu}^{{\rm WA}} - a_{\mu}^{{\rm SM}} = 251(59)\times 10^{-11}~.
\label{eq:aR}
\end{equation}

 The SM prediction includes the electroweak, QED and hadronic components contributing to the  muon $g-2$. Despite large uncertainties arising from $\alpha_{S}(M_{Z})$ recent studies have significantly improved the calculations of the hadronic contributions \cite{Cvetic:2020unz,Chakraborty:2018iyb,Izubuchi:2018tdd}. The leading order (LO) contribution from the hadronic vacuum polarization (HVP) yields  $a^{\rm HVP-LO,lattice}_\mu=707(55)\times10^{-11}$ ~\cite{Borsanyi:2020mff} using Lattice QCD, and comparison with the $e^{+}e^{-}\rightarrow ~ {\rm hadrons}$ data indicates a deviation of about $1.6\sigma$ between the experimental measurements and theoretical calculations of the hadronic contributions to the muon $g-2$ \cite{Davier:2019can,Tanabashi:2018oca}. This discrepancy in the hadronic sector can be ameliorated by reducing the overall uncertainty in the hadronic sector  such that $a_{\mu}^{{\rm HVP-LO,lattice}}$ may not need to require  extra contributions from strongly interacting new particles within BSM models to be consistent with the experimental measurements. On the other hand, this leaves the muon $g-2$ anomaly as expressed in Eq.(\ref{eq:aR}) to be explained by the QED and electroweak sectors. The discrepancy between theory and experiment remains even after the calculation is performed with high precision, and thus it indicates the need for contributions from QED and electroweak sectors of BSM models. A nice compendium of experimental and theoretical results is found in \cite{muoninitiative}.

While the muon $g-2$ anomaly can be used to constrain the electroweak sector of BSM models, another restriction is introduced by the precise experimental measurements of the electron $g-2$ as $a_{e}^{{\rm exp}} =1.15965218076(28) \times 10^{-3} $ \cite{Aoyama:2014sxa}. When the BSM models are built in a flavor blind fashion, their $g-2$ predictions for the electron and muon are correlated by $\Delta a_{e}/\Delta a_{\mu} = m_{e}^{2}/m_{\mu}^{2}$. The SM prediction calculated up to ten loops within QED \cite{Volkov:2018jhy,Aoyama:2017uqe,Volkov:2017xaq,Aoyama:2019ryr} yields $a_e^{\rm SM}=1.159652181643(25)(23)(16)(763) \times 10^{-3}$, which is, in contrast to the muon $g-2$ prediction, greater than the experimental results for the electron $g-2$,
\begin{equation}
\Delta a_{e} = -(8.8 \pm 3.6)\times 10^{-13}~.
\label{eq:Dae}
\end{equation}

Such a discrepancy does not lead to a numerical deviation only, but it also indicates that a mechanism must be implemented in BSM models which breaks lepton universality of the SM so that  $\Delta a_{e}/\Delta a_{\mu} = (-14)m_{e}^{2}/m_{\mu}^{2}$ as experimentally established. Numerous studies resolving the muon $g-2$ anomaly exist within a variety of BSM models \cite{Ferreira:2021gke,Arcadi:2021cwg,Zhu:2021vlz,Bai:2021bau,Das:2021zea,Ge:2021cjz,Brdar:2021pla,Buen-Abad:2021fwq,Wang:2021fkn,Li:2021poy,Calibbi:2021qto,Athron:2021iuf,Claude:2021sye,Frank:2020smf,Dasgupta:2021dnl,Balkin:2021rvh, Ahmed:2021htr,Abdughani:2021pdc,VanBeekveld:2021tgn,Cox:2021gqq,Endo:2021zal,Wang:2021bcx,Gu:2021mjd,Cao:2021tuh,Yin:2021mls,Han:2021ify,Aboubrahim:2021rwz,Yang:2021duj,Baum:2021qzx,Baer:2021aax,Altin:2017sxx,Frank:2017ohg,Frank:2020kvp,Gogoladze:2016jvm,Chakraborti:2021bmv,
Babu:2021jnu,Crivellin:2021rbq,ColuccioLeskow:2016dox,Hiller:2019mou,Crivellin:2020tsz,Shafi:2021jcg,Kowalska:2020zve,Kowalska:2017iqv}, while there also are (fewer) studies which attempt to provide a consistent description of both deviations \cite{Escribano:2021css,Cadeddu:2021dqx,Han:2021gfu,Cao:2021lmj,Li:2020dbg,Jana:2020pxx,Jana:2020joi,Banerjee:2020zvi}.

Given some attractive features such as the resolution to the gauge hierarchy problem, the existence of dark matter and a rich spectrum  testable at the ongoing experiments, supersymmetry (SUSY) and supersymmetric extensions of SM (SSM)
provide a class of well-motivated BSM models to explore the $g-2$ implications. In the minimal SSM (MSSM) framework, as is well known, the chirality flip between the left-handed and right-handed sleptons enhances the supersymmetric contributions to the leptons $g-2$  \cite{Martin:1997ns}. Since this supersymmetric enhancement is proportional to the lepton mass and $\mu\tan\beta$, the flavour blind interactions can adjust the magnitudes of the $g-2$ calculations but still yield the ratio $\Delta a_{e}/\Delta a_{\mu} = m_{e}^{2}/m_{\mu}^{2}$. However one can modify this ratio by choosing opposite signs for the bino and wino masses, in which the bino can provide the  main contribution to the electron $g-2$, while the Wino contribution is mainly responsible for the SUSY contributions to the muon $g-2$. However resolutions to the $g-2$ anomalies can simultaneously be accommodated when the chargino, neutralino and sleptons are very light ($\sim \mathcal{O}(100)$ GeV) \cite{Badziak:2019gaf}. The collider experiments are expected to impose strong constraints on such solutions, especially during the LHC-Run3 \cite{Fiaschi:2019zgh}. 

Since the gauge interactions are flavor universal, and Higgs/higgsino couplings to the electron and muon are quite small to significantly modify the  the SM predictions, it can be concluded that MSSM can adjust the $g-2$ of the electron and muon with the experimental data only with the light sleptons and electroweakinos. On the other hand,  SUSY models extending the MSSM with new particles and/or symmetries can potentially accommodate the simultaneous resolutions to the electron and muon $g-2$ anomalies by introducing new interactions involving non-MSSM fields, which can distinguish between the electron and muon. The simplest class of extended MSSM models can be built by supplementing the MSSM gauge symmetry with an extra Abelian $U(1)^\prime$ gauge group \cite{Cvetic:1996mf,Hewett:1988xc,Cvetic:1995rj}. This class of models introduces a new neutral gauge boson ($Z^\prime$) associated with the local $U(1)^\prime$ symmetry. Besides, one can implement a spontaneous breaking mechanism for $U(1)^\prime$ which requires another field ($S$). The $S$ field is preferably a singlet under the MSSM gauge symmetry, while it is non-trivially charged under the $U(1)^\prime$ symmetry so that its vacuum expectation value (VEV) can break the $U(1)^\prime$ symmetry, while leaving the MSSM symmetry intact. In addition, the supersymmetric partners of these new particles are also included in the spectrum and,  since they are expected to interfere with the MSSM particles, they can significantly alter the low scale implications of the model. In addition to these particles, such models need three MSSM singlet fields (one for each matter family) to cancel the gauge and gravity anomalies. These singlet fields can be sometimes chosen to be the right-handed neutrinos, and so they can provide a suitable framework for the neutrino masses and mixing.  Whereas in general $U(1)^\prime$ models one must either introduce exotic fermion fields, or add three scalar singlets (secluded $U(1)^\prime$ models) to cancel all anomalies.

This class of models have been  explored \cite{Hicyilmaz:2016kty,Hicyilmaz:2017nzo,Frank:2020byg,Frank:2020ixv,Frank:2020pui,Frank:2019nwk,Cincioglu:2010zz,Cleaver:1997nj} by assuming family-universal $U(1)^\prime$ charges for the MSSM fields inspired by the lepton universality of the SM. However, it is also possible to consider sets of $U(1)^\prime$ charges in which different families of matter fields can have distinct charges \cite{Duan:2018akc}. Even though it seems a simple difference and it does not change the spectrum compared to the $U(1)^\prime$ extended MSSM models with universal charges, the non-universal charges can change the particle interactions by forbidding some terms while allowing others which are not present in the case of universal $U(1)^\prime$ charges. This would provide a natural expression for the apparent non-universality of electron and muon magnetic moments. 

Motivated by its different structure we will explore the class of $U(1)$ extended MSSM models in which the matter families are charged differently under the $U(1)^\prime$ gauge group, whose salient features will be discussed in Section \ref{sec:model}. The rest of the paper is organized as follows.  After explaining our scanning procedure and enforcing experimental constraints in Section \ref{sec:WpZpmass}, we present our lepton anomalous magnetic moment results in the surviving parameter space from LHC direct SUSY constraints and dark matter constraints  in Section \ref{sec:LepG2}.  We then choose three different benchmark scenarios, highlighting different dark matter choices, and investigate the detectability of non-universal $U(1)^\prime$ scenarios at the LHC in Section \ref{sec:Collider} and indicate the best variables to fulfill this goal. We then summarise and conclude in Section \ref{sec:conclusion}.

\section{Non-Universal $U(1)^\prime$ Models}
\label{sec:model}

A known problem in supersymmetry is the so-called $\mu$ problem \cite{Cvetic:1996mf,Suematsu:1994qm,Langacker:1998tc}. That is, even though the $\mu-$parameter is directly related to the electroweak (EW) symmetry breaking, its scale can lie between the EW scale and the grand unification (or Planck) scale, since it is not protected by any symmetry. This is resolved if the MSSM gauge group is extended to involve more local gauge groups, and the MSSM fields are non-trivially charged under the extra symmetry groups, then the $\mu-$term can be generated dynamically and its scale can be restricted by the breaking of the symmetries. Indeed, supersymmetric $U(1)^\prime$ models were originally motivated by stabilizing the $\mu-$term at the scales consistent with the EW symmetry breaking. This class of supersymmetric models can be motivated by the SUSY grand unified theories (GUTs), since additional $U(1)$ groups can emerge from the breaking of grand unified groups larger than $SU(5)$ such as $SO(10)$ and $E_{6}$, which are extensively explored in \cite{Suematsu:1994qm,Hewett:1988xc,Langacker:1998tc, Frank:2020pui}. 

In addition to stabilizing the $\mu$-term,  the new particles and interactions required by the $U(1)'$ invariance and non-trivial charges of the MSSM fields under the $U(1)'$ gauge group also enrich the supersymmetric spectrum and phenomena. As summarized in the previous section, a local $U(1)^\prime$ group requires a neutral gauge boson  ($Z^\prime$) in the spectrum. Indeed, $Z^\prime$ has been under investigation experimentally in a model dependent framework for a while, and the current bounds on the $Z^\prime$ mass for $U(1)^\prime$ extended supersymmetric models is $M_{Z^\prime} \gtrsim 5$  TeV \cite{Aad:2019fac, Sirunyan:2021khd}. Even though the impact from this heavy new gauge boson can be suppressed in the experimental observations, this is not the case for the mass of its supersymmetric partner, $\tilde{B}^\prime$, which can even be as light as about 100 GeV \cite{Khalil:2015wua}. In addition,  the spectrum should include another superfield $\hat{S}$,  whose scalar component is responsible for the spontaneous  breaking of $U(1)^\prime$ by developing a non-zero VEV. This VEV is also responsible for generating the $\mu$-term dynamically. In addition to these new particles,  the anomaly cancellation conditions require more particles in the spectrum. The properties of new particles depend on the considered $U(1)^\prime$ group and charges under this symmetry. For instance, if $U(1)^\prime \equiv U(1)_{B-L}$, then the anomaly cancellation requires three fields which are singlet under the MSSM gauge group. One can naturally prefer to include right-handed neutrinos as one per each family to cancel the anomalies \cite{Un:2016hji,Khalil:2010iu}, but  right-handed neutrinos can cancel the anomalies {\it only} when the MSSM gauge group is extended by $U(1)_{B-L}$. In general, $U(1)^\prime$ models require also exotic superfields to cancel the anomalies \cite{Cheng:1998nb,Cheng:1998hc,Erler:2000wu,Langacker:2000ju,Barger:2003hg,Demir:2005ti}.  

\subsection{Family-Dependent Charge Assignments}
\label{subsec:charges}
\begin{table}[t!]
	\begin{center}
	\setlength\tabcolsep{15pt}
	\renewcommand{\arraystretch}{1.5}			
		\begin{tabular}{|c|c|c|c|c|c|}
			\hline \hline
			SF & Spin 0 & Spin \(\frac{1}{2}\) & Generations & $U(1)_Y\otimes\, SU(2)_L \otimes\, SU(3)_C\otimes\, U(1)^\prime$  \\
			\hline
			\(\hat{Q_i}\) & \(\tilde{Q_i}\) & \(Q_i\) & 3 & \((\frac{1}{6},{\bf 2},{\bf 3}, Q_{Q_i}) \) \\
			\(\hat{L}_i\) & \(\tilde{L}_i\) & \(l_i\) & 3 & \((-\frac{1}{2},{\bf 3},{\bf 1}, Q_{L_i}) \) \\			
			\(\hat{H}_d\) & \(H_d\) & \(\tilde{H}_d\) & 1 & \((-\frac{1}{2},{\bf 2},{\bf 1}, Q_{H_d}) \) \\
			\(\hat{H}_u\) & \(H_u\) & \(\tilde{H}_u\) & 1 & \((\frac{1}{2},{\bf 2},{\bf 1}, Q_{H_u}) \) \\
			\(\hat{D}_i^c\) & \(\tilde{D}_{Ri}^*\) & \(d_{Ri}^*\) & 3 & \((\frac{1}{3},{\bf 1},{\bf \overline{3}}, Q_{D_i^c}) \) \\
			\(\hat{U}_i^c\) & \(\tilde{U}_{Ri}^*\) & \(u_{Ri}^*\) & 3 & \((-\frac{2}{3},{\bf 1},{\bf \overline{3}}, Q_{U_i^c}) \) \\
			\(\hat{N}^{c}_i\) & \(\tilde{N}^c_i\) & \(N^c_i\) & 3 & \((1,{\bf 3},{\bf 1}, Q_{N_i^c}) \) \\
			\(\hat{E}^{c}_i\) & \(\tilde{E}_{i R}^{*}\) & \(E_{i R}^{*}\) & 3 & \((1,{\bf 1},{\bf 1}, Q_{E^c_i}) \) \\			
			\(\hat{S}\) & \(S\) & \(\tilde{S}\) & 1 & \((0,{\bf 1},{\bf 1}, Q_S) \) \\	
			\hline \hline
		\end{tabular}
		\caption{ \label{tab:superfields} Superfield configuration in the non-universal $U(1)^\prime$ model.}
	\end{center}
\end{table}

In Table \ref{tab:superfields} we list the all the particles involved in the spectrum and their charges under the MSSM gauge group and the additional $U(1)^\prime$ symmetry. 
The Higgs VEVs responsible for breaking the $U(1)^\prime$ model to $U(1)_{\rm em}$ are
\begin{eqnarray}
\langle H_u \rangle = \frac{1}{\sqrt{2}} \left(\begin{array}{c}
0 \\
v_u  \\ 
\end{array}\right) \, \qquad \langle H_d \rangle = \frac{1}{\sqrt{2}} \left(\begin{array}{c}
v_d \\
0  \\ 
\end{array}\right) \, , \qquad  \langle S \rangle =\frac{1}{\sqrt{2}} v_S\, ,
\label{eq:higgsfield}
\end{eqnarray}
with $\tan \beta=\frac{v_u}{v_d}$.
In Table \ref{tab:superfields} the subscript $i$ runs over the families. We assume family-dependent charges for the leptons, while the quarks carry family-independent charges{\footnote{We discuss this requirement further at the end of this section.}. We display the $U(1)^\prime$ charges as variables in the table, since one can configure different sets for the charges. These sets can be determined by imposing  gauge invariance and  anomaly cancellation conditions. The gauge invariance requires the following equations satisfied simultaneously:

\begin{equation}
\setstretch{1.5}
\begin{array}{l}
\left. \begin{array}{ll}
Q_{Q_{i}}+Q_{U^{c}_{i}}+Q_{H_{u}} & = 0~,  \\
Q_{Q_{i}}+Q_{D^{c}_{i}}+Q_{H_{d}} & = 0 ~,\\
Q_{L_{3}}+Q_{E^{c}_{3}}+Q_{H_{d}} & = 0~, \\
Q_{L_{3}}+Q_{N^{c}_{3}}+Q_{H_{u}} & = 0~, \\
Q_{S}+Q_{H_{u}}+Q_{H_{d}} & = 0~,
\end{array}\right \rbrace\hspace{0.3cm}{\rm Superpotential} \\
\left.\begin{array}{ll}
Q_{L_{1}}+Q_{E^{c}_{1}}+Q_{H_{u}} & = 0~, \\
Q_{L_{2}}+Q_{E^{c}_{2}}+Q_{H_{u}} & = 0~.
\end{array}\right\rbrace \hspace{0.3cm}{\rm Non-Holomorphic~ Terms}
\end{array}
\label{eq:gaugeinvarianceset}
\end{equation}

In addition, for the model to be anomaly-free the $U(1)^\prime$ charges of chiral fields must satisfy conditions corresponding to the
vanishing of $U(1)^{\prime}$-$SU(3)$-$SU(3)$, $U(1)^{\prime}$-$SU(2)$-$SU(2)$,
$U(1)^{\prime}$-$U(1)_Y$-$U(1)_Y$, $U(1)^{\prime}$-graviton-graviton,
$U(1)^{\prime}$-$U(1)^{\prime}$-$U(1)_Y$ and $U(1)^{\prime}$-$U(1)^{\prime}$-$U(1)^\prime$ anomalies. 
An anomaly free $U(1)^\prime$ model should provide a solution for all the charges. 
However, if one assigns family-independent charges for the quark fields, as we did, then these equations cannot be solved simultaneously without inclusion of a number of exotic fields. If we require family-universal $U(1)^\prime$ charges for all quarks, this necessitates introducing a set of $n_{D_x}$ exotic fields ($\widehat{D}_{x}$, $\widehat{\bar{D}}_{x}$), whose charges can be determined with respect to $Q_{S}$, as necessary to solve all the equations for  gauge invariance and anomaly cancelation conditions.
The gauge conditions are then augmented by:
\begin{equation}
Q_{S}+Q_{D_x}+Q_{{\bar D}_x}  = 0
\end{equation}
while the anomaly cancellation conditions become
\begin{eqnarray}
0=&&\sum_i(2Q_{Q_i}+Q_{U^c_i}+Q_{D_i})+n_{D_x}(Q_{D_x}+Q_{{\bar D}_x}) \nonumber \\
0=&&\sum_i(3Q_{Q_i}+Q_{L_i})+Q_{H_d}+Q_{H_u} \nonumber \\
0=&&\sum_i(\frac{1}{6}Q_{Q_i}+\frac{1}{3}Q_{D^c_i}+
 \frac{4}{3}Q_{U^c_i}+\frac{1}{2}Q_{L_i}+Q_{E^c_i})+
 \frac{1}{2}(Q_{H_d}+Q_{H_u})
+3 n_{D_x}Y_{D_x}^2(Q_{D_x}+Q_{{\bar D}_x}\!) \nonumber \\
0=&&\sum_i(6Q_{Q_i}+3Q_{U^c_i}+3Q_{D^c_i}+2Q_{L_i}+Q_{E^c_i})+2Q_{H_D}
 +2Q_{H_u}+Q_s
 +3 n_{D_x}(Q_{D_x}+Q_{{\bar D}_x}\!)\nonumber \\ 
0=&&\sum_i(Q_{Q_i}^2+Q_{D^c_i}^2-2Q_{U^c_i}^2-Q_{L_i}^2+Q_{E^c_i}^2)-
 Q_{H_d}^2+Q_{H_u}^2 +3 n_{D_x}Y_{D_x} (Q^2_{D_x}-Q^2_{{\bar D}_x})\nonumber \\
0=&&\!\!\!\sum_i(6Q_{Q_i}^3+3Q_{D^c_i}^3+3Q_{U^c_i}^3+2Q_{L_i}^3+Q_{E_i}^3)+
 2Q_{H_d}^3+2Q_{H_u}^3+Q_S^3
 +3n_{D_x}(Q^3_{D_x}+Q^3_{{\bar D}_x}\!)
 \label{eq:anomalies2}
\end{eqnarray}
where $n_{D_x}$ are the number of exotic fields,  $Y_{D_x}$ is their hypercharge and $Q_{D_x},\,  Q_{{\bar D}_x}$ their $U(1)^\prime$ charges. All equations are satisfied by setting $n_{D_x}=3$, $Y_{D_x}=-1/3$, while for $Q_{D_x}, Q_{{\bar D}_x}$ several values are possible depending on the choice of $U(1)^\prime$ charges for the other particles. These fields which behave effectively as exotic $d$-type quarks, can be thought of as remnants from the breaking of $E_6$.  Their properties and consequences at LHC, have been considered in  \cite{Hicyilmaz:2021oyd}. They do not affect the phenomenology we study here, so we omitted them in Table \ref{tab:superfields} and we ignore them in our further considerations.

We proceed by giving, in the next subsection, the Lagrangian and soft-breaking terms which satisfy all gauge and anomaly canceling conditions.

\subsection{Lagrangian and particle masses}
\label{subsec:superpotential}

If the presence of exotic fields the particle spectrum is required by anomaly cancellation conditions, one can minimize the required number of exotic fields by assigning family-dependent $U(1)^\prime$ charges to the MSSM fields. Since our work focuses on $g-2$ of the electron and muon, we implement different charges to each lepton family only, and assume that the quarks have family-universal charges. However, family-dependent charges forbid leptons to couple to the Higgs field in the superpotential, which results in massless fermions \cite{Demir:2005ti}. If we assign a charge to the third lepton family which allows it to couple to the Higgs field, then the superpotential does not involve any Yukawa term for the electron and muon. We make this choice as the $\tau$ is heavier and its mass significantly different from zero. In this case, the superpotential can be written as follows:
\begin{eqnarray}
\!\!\!\!\!\!	\widehat{W}^\prime_{\rm non-Uni}\!&=&Y_u\widehat{Q}\cdot \!\widehat{H}_u \widehat{U}+
Y_d\widehat{Q}\cdot \! \widehat{H}_d \widehat{D} + Y_{\tau}\widehat{L}_3\cdot \!
\widehat{H}_d \widehat{E}_3 + \lambda \widehat{S}\widehat{H}_u \cdot \!
\widehat{H}_d + {Y_{\nu}} \widehat{L}\cdot \!
\widehat{H}_u \widehat{N} + {\kappa} \widehat{S} \widehat{D}_x
\widehat{{\overline{D}}}_x
\label{eq:superpot}
\end{eqnarray}
Thus our consideration of the family-dependent charges  in the leptonic sector still requires introducing only two exotic fields, $\widehat{D}_{x}$ and $\widehat{\bar{D}}_{x}$ coupling to the MSSM singlet $\widehat{S}$ field, see \cite{Hicyilmaz:2021oyd}, where this was shown to be a non-universal $U(1)^\prime$ scenario with a minimal extension. Here $\widehat{Q}$ denotes the left-handed quark superfields, and $\widehat{U}, \widehat{D}$ represent the right-handed up-type and down-type quark fields, respectively. As we stated above, the family-dependent charges for the lepton superfields allow only the $\tau$ superfield ($\widehat{L}_{3}$) to interact with $H_{d}$, in our choice, while the Yukawa terms for the first two-family leptons are forbidden. In addition, since the MSSM Higgs fields, $\widehat{H}_{u}$ and $\widehat{H}_{d}$ are non-trivially charged under the $U(1)^\prime$ group, their bilinear mixing is also forbidden. Instead, the superpotential involves $\lambda \widehat{S}\widehat{H}_{u}\widehat{H}_{d}$. The bilinear mixing of these Higgs fields is effectively generated by the VEV of $S$ as $\mu_{{\rm eff}}=\lambda \langle S \rangle$. Finally,  we introduce right-handed neutrinos, and allow a Yukawa term as $Y_{\nu} \widehat{L}\widehat{H}_u \widehat{N}$. 
We expect the right-handed neutrinos to decouple and to not affect directly the low scale  implications{\footnote{However their superpartners, the sneutrinos, could be relevant, and in particular one of them could be the lightest supersymmetric particle (LSP) and thus a candidate for dark matter. We comment on this choice later.}.

Based on the superpotential given in Eq.(\ref{eq:superpot}), the soft supersymmetry breaking (SSB) can be achieved through the following Lagrangian:

\begin{align}
-{\mathcal{L}}^\prime_{\rm soft} &= \sum_i M_i \lambda_i\lambda_i-A_{\lambda}\lambda S H_dH_u 
-A_u Y_u U^c Q H_u-A_{d} Y_d D^c QH_d-
A_\tau Y_\tau E_3 L_3 H_d-A_{\kappa}\kappa S {D_x {\overline{D}_x}} +h.c.  \nonumber \\
&+ m_{H_u}^2|H_u|^2+ m_{H_d}^2|H_d|^2+m_S^2|S|^2+ 
m_{Q}^2 \widetilde Q\widetilde Q
+m_{U}^2 \widetilde U^c\widetilde U^{c}+m_{D}^2 \widetilde D^c\widetilde D^{c}
+m_{L}^2 \widetilde L\widetilde L\nonumber \\
&+ m_{E}^2 \widetilde E^c\widetilde E^{c}+m_{X}^2 \widetilde{D}_x \widetilde{D}_x+m_{\bar{X}}^2 \widetilde{\overline{D}}_x \widetilde{\overline{D}}_x +h.c. \, ,
\label{eq:soft}
\end{align}
where $m_{\tilde{Q}}$, $m_{\tilde{U}}$, $m_{\tilde{D}}$, $m_{\tilde{E}}$, $m_{\tilde{L}}$, $m_{H_{u}}$, $m_{H_{d}}$, $m_{S}$, $m_{X}$  and $m_{\bar{X}}$ are the mass matrices of the scalar particles,  $M_{i}\equiv M_{1},M_{2},M_{3},M_{4}$ are  gaugino masses, and  $A_{\lambda}$, $A_{u}$, $A_{d}$ , $A_{\tau}$ and $A_{\kappa}$ are the trilinear scalar interaction couplings. 

Within these interactions the model still suffers from having massless leptons for the first two families, inconsistent with experiment, as the absence of Yukawa terms for these leptons prevents them to acquire masses at tree-level. Moreover, there is no term which can yield masses for the leptons at loop-level. One must then consider non-holomorphic terms \cite{Demir:2005ti}, where the superpartners of these massless leptons couple to the ``wrong" Higgs doublet. Even though these interaction terms do not involve the SM leptons, they can induce masses for the leptons at loop-level. The following terms are necessary to complete the Lagrangian which then yields consistent masses for all the SM particles:
\begin{equation}
-\mathcal{L}^\prime_{\rm non-holomorphic}=T_{e}^\prime \tilde{L}_1 H_{u}^*\tilde{E}_1^{c} +T_{\mu}^\prime \tilde{L}_2 H_{u}^*\tilde{E_2^{c}}+h.c. \, ,
\label{UMSSM_NH}
\end{equation}  
where $T_{e}^\prime$ and $ T_{\mu}^\prime$  are the corresponding non-holomorphic couplings for electrons and muons. We then can proceed  by adopting a non-universal $U(1)^\prime$ model with a minimum number of exotics from \cite{Hicyilmaz:2021oyd}. The non-holomorphic terms introduced through Eq.(\ref{UMSSM_NH})  induce the following effective mass terms for the leptons at one-loop involving the neutralino exchange \cite{Borzumati:1999sp}.
\begin{equation}
m_{f} = T_{i}^\prime v_{u}\dfrac{g_{1}^{2}}{8\pi^{2}}\displaystyle \sum_{j=1}^{6} K_{f}^{j}m_{\tilde{\chi}_{j}^{0}}I\left(m_{\tilde{f}_{1}}^{2},m_{\tilde{f}_{2}}^{2},m_{\tilde{\chi}_{j}^{0}}^{2} \right)~,
\label{eq:loopcont}
\end{equation}
where $T_{i}^\prime$ with $i=e,\mu$ are the non-holomorphic trilinear couplings between the sleptons and $H_{u}$, $K_{f}^{i}$ is the coupling between the neutralino and leptons, and $I\left(m_{f_{1}}^{2},m_{\tilde{f}_{2}}^{2},m_{\tilde{\chi}_{j}^{0}}^{2} \right)$ stands for the loop function, which is given by 
\begin{equation}
\setstretch{3}
 I\left(m_{1}^{2},m_{2}^{2},m_{3}^{2} \right) = \left\lbrace \begin{array}{ll}
 \dfrac{1}{m_{1}^{2}-m_{2}^{2}}\left(\dfrac{m_{1}^{2}\log(m_{3}^{2}/m_{1}^{2})}{m_{3}^{2}-m_{1}^{2}} - \dfrac{m_{2}^{2}\log(m_{3}^{2}/m_{2}^{2})}{m_{3}^{2}-m_{2}^{2}}\right) &, \, {\rm no-degeneracy} \\
\dfrac{m^{2}}{(m_{3}^{2}-m^{2})^{2}}\left(1-\dfrac{m_{3}^{2}}{m^{2}}+\dfrac{m_{3}^{2}}{m^{2}}\log(m_{3}^{2}/m^{2}) \right) & ,\, m_{1}\sim m_{2} = m \\
\dfrac{\log(m_{3}^{2}/m_{2}^{2})}{m_{3}^{2}-m_{2}^{2}} &, \, m_{1} \ll m_{2}~.
 \end{array}\right.
 \label{eq:loopFunc}
 \end{equation} 

 Before concluding the discussion on the charge assignments we should also note that assigning family-dependent charges to the quark fields may remove the need for the exotic fields (see, for instance, \cite{Demir:2005ti}). In such cases Yukawa terms for the down-type quarks are also forbidden, and one can add other non-holomorphic terms to Eq.(\ref{UMSSM_NH}) which induce effective $m_{d_{i}}$ terms for the down-type quarks. The quarks could acquire their masses at one-loop level through the gluino exchange as well as those involving the neutralino. The gluino contribution to the quark masses at one-loop can be written as
\begin{equation}
m_{f} = T_{d_{i}}^\prime v_{u}\dfrac{g_{3}^{2}}{6\pi^{2}}\displaystyle m_{\tilde{g}}I\left(m_{\tilde{f}_{1}}^{2},m_{\tilde{f}_{2}}^{2},m_{\tilde{\chi}_{j}^{0}}^{2} \right)~,
\label{eq:glucont}
\end{equation}
 where $T_{d_{i}}^\prime$ are would-be non-holomorphic trilinear scalar couplings between the down-type squarks and $H_{u}$ similar to those given for leptons in Eq.(\ref{eq:loopcont}). Combining this result with the one from the loop function given in Eq.(\ref{eq:loopFunc}) the gluino contribution seems to be inversely proportional to the gluino mass, $m_{\tilde{g}}$. This dependence has a strong impact on the loop level quark masses due to the heavy mass bounds on the gluino and the squarks \cite{Aad:2019ftg,Vami:2019slp}. Even though one can consider large $T_{d}^\prime$ terms to enhance the gluino contribution, large $T_{d}^\prime$ causes a large mass difference ($m_{\tilde{f}_{2}}^{2}-m_{\tilde{f}_{1}}^{2}$) between the squarks, which provides another suppression in the loop function. In this case, the loop-level quark masses mostly depends on the contributions from the neutralino exchange given in Eq.(\ref{eq:loopcont}). The neutralino loops can provide consistent masses for the quarks from the first two families; however, a large mass of the bottom quark ($m_{b}\simeq 4.18$ GeV) requires significant contributions from the gluino loops, as well. Hence, the suppression from the heavy gluino and squarks effectively exclude the option in which the bottom quark acquires its mass through the loops (that is, the presence of non-holomorphic terms for down-type quarks). Considering this challenge, even though it is possible to realize charge assignments without a need for the exotics, we prefer a configuration in which all the quarks receive their masses at tree level, with the disadvantage of the inclusion of a set of exotic fields. 
 
 \vskip0.1in

\subsection{Anomalous Magnetic Moments--Theoretical Considerations}
\label{subsec:magmoment}
As discussed before, the hadronic contributions to the lepton $g-2$ within the SM are more or less consistent with the experimental measurements, and the main discrepancy arises from the electroweak contributions, which necessitate significant contributions from new physics. Typical electroweak contributions to the lepton $g-2$ in many supersymmetric models are provided by the slepton-neutralino and the chargino-sneutrino loops \cite{Moroi:1995yh,Martin:2001st,Giudice:2012pf}. These contributions are enhanced by the chirality flip between the sleptons or the mixing among the neutralino (chargino) species. If the masses of the particles running in loops are set to a common mass scale ($M_{{\rm SUSY}}$), the leading contributions within the MSSM framework can be  approximated by \cite{Athron:2021iuf}:

\begin{equation}
\Delta a_{l}\approx \dfrac{m_{l}^{2}}{m_{\mu}^{2}} C_{S}{\rm sign}(\mu M_{i})\left(\dfrac{500~{\rm GeV}}{M_{{\rm SUSY}}} \right)\dfrac{\tan\beta}{40}~, 
\label{eq:g2approx}
\end{equation}
where
\begin{equation*}
C_{S}=\left\lbrace \begin{array}{ll}
21\times 10^{-10} & {\rm for~WHL}~, \\
1.2\times 10^{-10} &  {\rm for~BHL}~, \\
-2.4\times 10^{-10} &  {\rm for~BHR}~, \\
\displaystyle 2.4 \left(\frac{\mu}{500~{\rm GeV}} \right) \times 10^{-10} &  {\rm for~BLR}~.
\end{array}\right.
\end{equation*}
Eq.(\ref{eq:g2approx}) can be used to quantify the contributions from different neutralino and chargino species. Following the notation of \cite{Athron:2021iuf} $B,W,H$ denote bino, wino and higgsino respectively, while $L,R$ represent the left and right-handed sleptons.  As seen from $C_{S}$, the BLR contribution is proportional to $\mu-$term which results from the chirality flip between the left- and right-handed sleptons, and so the dominant supersymmetric contribution  to lepton $g-2$ arises from bino-slepton loops. The enhancement from the chirality flip is proportional to $\mu\tan\beta -A_{l}$. Even though $A_{l}$ can be neglected  by comparison to $\mu\tan\beta$ for moderate and large values of $\tan\beta$, it can slightly break the relation between the electron and muon $g-2$ given as $\Delta a_{e}/\Delta a_{\mu} = m_{e}^{2}/m_{\mu}^{2}$, so that  contributions to $g-2$ can have different signs for electrons and muons when the sleptons are significantly light \cite{Baum:2021qzx}. 

The contributions from the class of $U(1)'$ supersymmetric models, in general, differ from the MSSM contributions by the inclusion of two additional neutralino species; i.e. $\tilde{B}^\prime$ and $\tilde{S}$ and their contributions. After the $U(1)^\prime$ and electroweak symmetry breaking, the  mass matrix  for the neutralinos in the $(\tilde{B}^\prime,\tilde{B},\tilde{W},\tilde{H}_{u},\tilde{H}_{d},\tilde{S})$ basis is:

\begin{equation}
	\mathcal{M}_{\tilde{\chi}^{0}}=\left(\begin{array}{cccccc}
	M_{4} & 0 & 0 & g^\prime Q_{H_{d}}v_{d} & g^\prime Q_{H_{u}}v_{u} & g^\prime Q_{S}v_S \\ 
	0 &M_{1}& 0&-\dfrac{1}{\sqrt{2}}g_{1}v_{d} & \dfrac{1}{\sqrt{2}}g_{1}v_{u}& 0 \\ 
	0 & 0&M_{2}&\dfrac{1}{\sqrt{2}}g_{2}v_{d}& -\dfrac{1}{\sqrt{2}}g_{2}v_{u}& 0 \\
	g^\prime Q_{H_{d}}v_{d} & -\dfrac{1}{\sqrt{2}}g_{1}v_{d}&\dfrac{1}{\sqrt{2}}g_{2}v_{d}&0&-\dfrac{1}{\sqrt{2}}\lambda v_{S}& -\dfrac{1}{\sqrt{2}} \lambda v_{u} \\ 
	g^\prime Q_{H_{u}}v_{u} &\dfrac{1}{\sqrt{2}}g_{1}v_{u}&-\dfrac{1}{\sqrt{2}}g_{2}v_{u}&-\dfrac{1}{\sqrt{2}} \lambda v_S&0& -\dfrac{1}{\sqrt{2}}\lambda v_{d} \\ 
	g^\prime Q_{S}v_S &  0 & 0 & -\dfrac{1}{\sqrt{2}} \lambda v_{u} & -\dfrac{1}{\sqrt{2}}\lambda v_{d} & 0
    \end{array}
	\right),
	\label{eq:neutralino_mass}
\end{equation}
which yields the neutralino mass eigenstates after diagonalization by a unitary matrix $N$ as
\begin{equation} 
N^* m_{\widetilde{\chi}^0} N^{\dagger} = m^{D}_{\widetilde{\chi}^0} ~.
\end{equation} 

Since the MSSM fields are non-trivially charged under the additional $U(1)^\prime$ group, the $\tilde{B}^\prime$ neutralino also participates in the processes which contribute to the lepton $g-2$. The $\tilde{B}^\prime$ contribution can be obtained through a loop process which can be identified as B$^{\prime}$LR, B$^{\prime}$HL and B$^{\prime}$HR in notation of \cite{Athron:2021iuf}. The contribution from all neutralinos to $g-2$ is obtained from the general expression of the supersymmetric $g-2$ contributions through neutralino-slepton loops in terms of neutralino mass eigenstates
\begin{eqnarray}
a_{l}^{\tilde{\chi}^0}=&&-\frac{m_{l}}{16\pi^2}\sum_{i=1}^{6}\sum_{j=1}^{2}\bigg[\big(|n^L_{ij}|^2+|n^R_{ij}|^2\big)\frac{m_{l}}{12{m_{\tilde{l}}^2}_j}F_1^N(x_{ij})
+\frac{{m_{\tilde{\chi}^0}}_i}{3{m_{\tilde{l}}^2}_j}\Re\big(n^L_{ij}n^R_{ij}\big)F_2^N(x_{ij})\bigg]
\end{eqnarray}
where $x_{ij}\equiv m_{\tilde{\chi}_{i}}^{2}/m_{\tilde{l}_{j}}^{2}$ and the functions in the loop are:
\begin{eqnarray}
F_1^N(x)&=&\frac{2(1-6x+3x^2+2x^3-6x^2\log x)}{(1-x)^4}~,\nonumber\\
F_2^N(x)&=&\frac{3(1-x^2+2x\log x)}{(1-x)^3}~.
\end{eqnarray}
Here $n^L_{ij}$ ($n^R_{ij}$) represent couplings of the left-handed (right-handed) leptons to the neutralinos given as follows:
\begin{eqnarray}
n^L_{ij}= && -Y_{L} N^{\star}_{i,3} X^{l \star}_{j,1} - \sqrt {2} g_1 N^{\star}_{i,1} X^{l \star}_{j,2} + \sqrt {2} g^\prime N^{\star}_{i,5} X^{l \star}_{j,2} \nonumber \\
n^R_{ij}=&&-Y_{L} N_{i,3} X^{l\star}_{j,2} + \big(\frac{g_1}{\sqrt {2}} N_{i,1}+ \frac{g_2}{\sqrt {2} } N_{i,2}- \sqrt {2} g^\prime N_{i,5}\big)X^{l\star}_{j,1}.\nonumber
\end{eqnarray}

The contribution from $\tilde{B}^\prime$ is associated with the coupling $g^\prime$ and the $U(1)^\prime$ charges of the leptons ($X^{l}$), and the suppression from the $\tilde{B}^\prime$ mass is controlled by the loop functions given above. Note that there is no contribution from $\tilde{S}$, since it does not couple to the leptons directly.

Despite the additional contributions from $B^{\prime}$, the correlation between the $\Delta a_{e}$ and $\Delta a_{\mu}$ still holds in the case of family-independent $U(1)^\prime$ charges, since lepton families are not distinguished. But a differentiation between the leptons can be realized by introducing family-dependent $U(1)^\prime$ charges, which is what motivates our study. Non-universal charges for electrons and muons under $U(1)^\prime$ mean that $\Delta a_{e}$ and $\Delta a_{\mu}$ can become almost independent on each other. Besides, if the mass parameter of $\tilde{B}^{\prime}$ is set to be negative ($M_{4}< 0$), then electron and muon $g-2$ can receive contributions with opposite signs, as desired in our work. 

In addition to the family-dependent charges, our model significantly differ from the general $U(1)^\prime$ models because of the presence of the non-holomorphic terms given in Eq.(\ref{UMSSM_NH}). Since the selectron and smuon couple to the ``wrong" Higgs field, the chirality flip between the left-handed and right-handed states of these sleptons becomes proportional to $\mu\cot\beta - A^{\prime}_{l}$, and consequently the $T^{\prime}_{l}-$term ($T^{\prime}_{l}\equiv A^{\prime}_{l}y_{l}$) dominates over the $\mu-$term due to the suppression from $\tan\beta$. In this case, setting opposite sign $T-$terms for selectron and smuon can easily yield $\Delta a_{e}$ and $\Delta a_{\mu}$ predictions, which can satisfy experimentally established correlation ($\Delta a_{e}/\Delta a_{\mu} = -14 m_{e}^{2}/m_{\mu}^{2}$) without the need for very light sleptons.

The additional $U(1)^\prime$ symmetry leaves the charged sector of MSSM intact, and thus the contributions from the chargino-sneutrino loops are the same as those realized in the MSSM framework{\footnote{With the understanding that the $\mu$-term is generated dynamically, and is proportional to $v_S$.}. Its general expression can be written as

\begin{eqnarray}
a_{l}^{{\tilde{\chi}}^{\pm}}=&&\frac{m_{l}}{16\pi^2}\sum_{k=1}^2\bigg[\frac{m_{l}}{12m_{\tilde{\nu}_l}^2}\big(|c^{L}_{k}|^2
+|c^{R}_{k}|^2\big)F_1^C(y^l_{k})
 +\frac{2m_{{\tilde{\chi}}_k^{\pm}}}{3m_{\tilde{\nu}_l}^2}\Re\big(c^{L}_{k}c^{R}_{k}\big)F_2^C(y^l_{k})\bigg]
\end{eqnarray}
where the loop functions are
\begin{eqnarray}
F_1^C(x)=&&\frac{2(2+3x-6x^2+x^3+6x\log x)}{(1-x)^4}\nonumber \\
F_2^C(x)=&&-\frac{3(3-4x+x^2+2\log x)}{2(1-x)^3}
\end{eqnarray}
and $ y^l_{k}=\frac{m_{{\tilde{\chi}}_k^{\pm}}^2}{m_{\tilde{\nu}_l}^2}$. The couplings between the leptons and the charginos are $c^{L}_{k}=y_{l}U^{\ast}_{k2}, \hspace{0.2cm}c^{R}_{k}=-g_2 V_{k1}$.  We should note that the presence of the additional $U(1)$ group is still effective in this sector, since the Higgsino masses (i.e. $\mu-$term) is dynamically generated through the breaking of $U(1)^\prime$ symmetry ($\mu_{{\rm eff}}\equiv\frac{1}{\sqrt{2}}  \lambda v_{S}$, where $v_{S}$ is the breaking scale if $U(1)^\prime$ symmetry.)

Before concluding we should also include the contributions to $\Delta a_{l}$ from $Z^\prime$ boson as \cite{Baek:2001kca,Ma:2001md,Heeck:2011wj}

\begin{equation}
\Delta a_{\mu}^{Z^{\prime}} = \frac{g^{\prime 2}m_{\mu}^2}{4\pi^2}\int_0^1 dz \frac{z^2(1-z)}{m_{\mu}^2z+M_{Z^{\prime}}^2(1-z)}.  
\end{equation}
This contribution to $(g-2)_{\mu}$ is always positive; however, it is quite suppressed due to the strict bound on the $Z^\prime$ as $M_{Z^\prime} > 5.5$ TeV.

\section{Computational setup and  experimental constraints}
\label{sec:WpZpmass}

Following the development of the model as in Sec. \ref{sec:model}, to enable our analysis and impose constraints coming from  experimental data, we implement the model within a computational framework. We used  \sa~ (version 4.14.3) \cite{Staub:2008uz,Staub:2010jh,Staub:2015kfa} to generate a UFO \cite{Degrande:2011ua} version of the model \cite{Christensen:2009jx} and {\textsc CalcHep} \cite{Belyaev:2012qa} model files, so that we could employ \mo~ (version 5.0.9) \cite{Belanger:2018ccd} for the computation of the predictions relevant for our dark matter study, \spheno~ (version 4.0.4) \cite{Porod:2003um,Porod:2011nf} package for spectrum analysis and \mg~ (version 3.0.3) \cite{Alwall:2014hca} for generating the hard-scattering event samples necessary for our collider study. We used  \ma~ \cite{Conte:2012fm} (version 1.8.58) for the analysis of section \ref{sec:Collider}. Note that  \sa~ (version 4.14.3) includes all RGE corrections to model parameters to second order, and these are intrinsically dependent on our choice of $U(1)^\prime$ charges. 

In addition, we have used the {\textsc PySLHA} package \cite{Buckley:2013jua} to read the input values for the  model  parameters  that  we  encode  under  the  {\textsc SLHA}  format \cite{Skands:2003cj}, and  to  integrate the various employed programmes into a single framework.  Using our interfacing, we performed a random scan of the model parameter space described in Table \ref{tab:scan_lim}  following the Metropolis-Hastings technique. Here $M_1$, $M_2$ and $M_3$ denote mass terms for MSSM gauginos while $M_4$ refers to the gaugino mass associated with the $U(1)^\prime$ gauge group. As before, $\tan\beta$ is the ratio of VEVs of the MSSM Higgs doublets. $\lambda$ is the coupling associated with the interaction of the $H_u$, $H_d$ and S fields. Trilinear coupling for $\lambda$ is defined as $A_\lambda \lambda$ at the SUSY scale. Note that, we also scan the Yukawa coupling, $Y_{ij}^\nu$, of the term $L_i H_u  N^c_j$ and we vary only the diagonal elements in the range of $1\times 10^{-8} - 1\times 10^{-7}$ while setting the off-diagonal elements to zero. Finally, the diagonal elements of slepton soft masses, $M_{ij}^\ell$, $M_{ij}^e$ and $M_{ij}^\nu$ are also varied between $1\times 10^{5} - 3\times 10^{7}$.

\begin{table}
	\centering
	\setlength\tabcolsep{8pt}
	\renewcommand{\arraystretch}{1.3}
	\begin{tabular}{cc|cc}
		Parameter  & Scanned range & Parameter      & Scanned range \\
		\hline
		$\tan\beta$ & [2., 60.]   & $v_S$    &  [10., 25.] TeV\\
		$M_1$  & [-3., 3.] TeV          &  $M_3$ & [1., 60.] TeV\\
		$M_2$ & [-5., 5.] TeV            &  $M_4$ & [-4., 4.] TeV \\
		\hline
		$\lambda$ & [0.02, 0.5]       &   $T^\prime_e$ & [-10., 10.] TeV \\
		$A_\lambda$ & [-3., 15.] TeV      &$T^\prime_\mu$ & [-15., 15.] TeV \\
	\end{tabular}
	\caption{Scanning range of parameter space of the Non-Universal $U(1)^\prime$ model.
}
	\label{tab:scan_lim}
\end{table}
We begin by scanning the $U(1)^\prime$ charges and use the following to restrict our choices.
\begin{enumerate}
	\item With the $U(1)^\prime$ charge  choices for the model, we obtain 416 unique charge sets, which are consistent with gauge and anomaly cancellation conditions.
	\item Requiring non-zero charges for the following: $Q_Q$, $Q_{U^c}$, $Q_{D^c}$, $Q_{H_u}$, $Q_{H_d}$, only 170 distinct sets survive.
	\item If we assume, in addition to non-zero charge conditions, that the ratios  $Q_{E^c_1}/Q_{E^c_2} > 2$ and  $Q_{L_1}/Q_{L_2} > 2$, to enhance electron over muon magnetic moment, we are left with 10 unique sets.  We present the surviving ten sets of charges in Table \ref{tab:u1solutions}.
\end{enumerate}

\begin{table}{
\setlength\tabcolsep{2pt}
\renewcommand{\arraystretch}{1.6}
\begin{tabular}{r||r|r|r|r|r|r|r|r|r|r}
\small
\centering	
& {\bf S1} & {\bf S2} &{\bf S3} & {\bf S4} & {\bf S5} & {\bf S6} & {\bf S7} & {\bf S8} & {\bf S9} & {\bf S10}	\\
\hline\hline
$Q_u$ &  0.166667  & 0.142857 & 0.444444 &  0.4 & -0.3  &-0.4 & -0.444444 & -0.142857 & -0.166667 &  0.3
\\
$Q_d$ & 0.333333 &  0.285714 &  0.222222 &  0.2 & -0.2 & -0.2 & -0.222222 & -0.285714 &  -0.333333 &  0.2
\\
$Q_{E^c_1}$ & 0.333333 & 0.428571 &  -1.000000 & -1.0 & 1.0 & 1.0 &1.000000 & -0.428571 & -0.333333 & -1.0
 \\
$Q_{E^c_2}$ &  0.000000 & -0.142857 &  -0.444444 & -0.3 & 0.4 & 0.3 & 0.444444 & 0.142857 & 0.000000 & -0.4
\\
$Q_{E^c_3}$ & 0.166667 & 0.142857 & 0.777778 &  0.7 & 0.1 & -0.7 & -0.777778 & -0.142857 & -0.166667 & -0.1 
\\
$Q_Q$ & 0.166667 & 0.142857 & 0.111111 & 0.1 & 0.1 & -0.1 & -0.111111 & -0.142857 & -0.166667 & -0.1
\\
$Q_{L_1}$ & -0.666667 & -0.714286 & 0.444444 &0.5 & -0.8 & -0.5 & -0.444444 & 0.714286 & 0.666667 & 0.8
\\
$Q_{L_2}$ &    -0.333333 & -0.142857 &   -0.111111 &  -0.2 &  -0.2 & 0.2 & 0.111111 & 0.142857 & -0.333333 & 0.2
\\
$Q_{L_3}$ & -0.333333 & 0.285714 & -0.444444 &-0.4 & -0.2 & 0.4 & 0.444444 & -0.285714 & -0.333333 & 0.2 
\\      
$Q_{N^c_1}$ &     1.000000  & 1.000000  & 0.111111 & 0.0 & 0.6  & 0.0 &-0.111111 & -1.000000 & -1.000000 & -0.6 
\\
$Q_{N^c_2}$ &  0.666667 & 0.428571 & 0.166667 & 0.7  & 0.0  & -0.7  & -0.666667 & -0.428571 &  -0.666667 &  0.0
\\
$Q_{N^c_3}$ &     0.0    & 0.0    & 0.0    & 0.0    & 0.0    & 0.0    & 0.0    & 0.0    & 0.0    & 0.0      
\\
$Q_{H_u}$ &     -0.333333  &  -0.285714 & -0.555556 & -0.5 & 0.2 & 0.5 & 0.555556 & 0.285714 & 0.333333 & -0.2
\\
$Q_{H_d}$ &      -0.500000     & -0.428571 & -0.333333  & -0.3  & 0.1  & 0.3  & 0.333333  & 0.428571 & 0.500000 & -0.1
\\
$Q_S$  &   0.833333 & 0.714286 &  0.888889 &  0.8 & -0.3 & -0.8 & -0.888889  & -0.714286  & -0.833333 &   0.3
\\
$Q_{D_x}$ &    -0.333333& -0.285714 & -0.555556  & -0.5 &  0.2 &  0.5 & 0.555556 &  0.285714 &  0.333333 & -0.2
 \\
 $Q_{{\bar D}_x}$ & -0.500000 & -0.428571 & -0.333333 & -0.3 &  0.1 &  0.3 &  0.333333 &  0.428571&  0.500000 & -0.1
 \\
 \end{tabular}
\caption{\label{tab:u1solutions}  $U(1)^\prime$ charges, as defined in Table \ref{tab:superfields}, for the 10 solutions ({\bf S1} - {\bf S10}) satisfying all imposed conditions.}}
\end{table}

We perform the random scan using the 10 unique sets obtained by requiring  $Q_{E^c_1}/Q_{E^c_2} > 2$ and  $Q_{L_1}/Q_{L_2} > 2$.

\begin{table}{
		\setlength\tabcolsep{7pt}
		\renewcommand{\arraystretch}{1.6}
		\begin{tabular}{l|c|c||l|c|c}
			\small
			\centering			
			Observable & Constraints & Ref. & Observable & Constraints & Ref.\\
			\hline
			$m_{h_1} $ & $ [122,128] $ GeV                     & \cite{Chatrchyan:2012xdj} &
			$m_{\widetilde{t}_1} $                                    & $ \geqslant 730 $ GeV & \cite{Tanabashi:2018oca}\\
			$m_{\widetilde{g}} $                                       & $ > 1.75 $ TeV & \cite{Tanabashi:2018oca} &
			$ m_{\widetilde{\chi}_1^\pm} $                     & $ \geqslant 103.5 $ GeV & \cite{Tanabashi:2018oca} \\
			$m_{\widetilde{\tau}_1} $                               & $ \geqslant 105 $ GeV & \cite{Tanabashi:2018oca} & 
			$m_{\widetilde{b}_1} $                                    & $ \geqslant 222 $ GeV & \cite{Tanabashi:2018oca}\\
			$m_{\widetilde{q}} $                                       & $ \geqslant 1400 $ GeV & \cite{Tanabashi:2018oca} &
			$m_{\widetilde{\mu}_1} $                               & $ > 94 $ GeV & \cite{Tanabashi:2018oca} \\
			$m_{\widetilde{e}_1} $                                    & $ > 107 $ GeV & \cite{Tanabashi:2018oca} &
			$ \lvert \alpha_{Z Z^\prime} \rvert $                            & $< \mathcal{O}(10^{-3})$ & \cite{Erler:2009jh} \\
			$M_{Z^\prime}$                                       & $> 5$ TeV & \cite{Aad:2019fac, Sirunyan:2021khd} &
			BR$(B^0_s \to \mu^+\mu^-) $ & $[1.1,6.4] \times10^{-9}$  &
			\cite{Aaij:2012nna} \\ 
			$\displaystyle  \frac{{\rm BR}(B \to \tau\nu_\tau)}
			{{\rm BR}_{SM}(B \to \tau\nu_\tau)} $ & $  [0.15,2.41] $ &
			\cite{Asner:2010qj} &
			BR$(B^0 \to X_s \gamma) $ & $  [2.99,3.87]\times10^{-4} $ &
			\cite{Amhis:2012bh}\\
		\end{tabular}
		\caption{\label{tab:constraints}  Current experimental and theoretical bounds used to determine consistent solutions in our scans.}}
\end{table}

LHC collaborations have explored possible signals originating from extra neutral gauge bosons $Z^\prime$. The most stringent constraint 
is derived from high-mass dilepton resonance searches which exclude $Z^\prime$ mass  up to 4.5 TeV from data accumulated at 
$\sqrt{s}=13$ TeV with $139~{\rm fb}^{-1}$ luminosity \cite{Aad:2019fac, Sirunyan:2021khd}{\footnote{Model-dependent analyses were performed, and thus mass limits on the $Z^\prime$ boson vary slightly among models.}. Search for heavy particles decaying into a 
top-quark pair yielded an exclusion limit on the $Z^\prime$ mass ranging from 3.1 TeV to 3.6 TeV at $\sqrt{s}=13$ TeV with 
$36.1~{\rm fb}^{-1}$ luminosity \cite{Aaboud:2019roo}. Dijet resonance search limits on $Z^\prime$ yielded slightly weaker limits, 
$M_{Z^\prime} > 2.7$ TeV at $\sqrt{s}=13$ TeV with $36~{\rm fb}^{-1}$ luminosity \cite{Sirunyan:2018xlo, Sirunyan:2019vgj}. Since 
 the most stringent constraints on $M_{Z^\prime}$ are obtained from its leptonic decay modes, models with a  
leptophobic $Z^\prime$ \cite{Babu:1996vt,Suematsu:1998wm,Chiang:2014yva,Araz:2017wbp,Frank:2020kvp} are much less constrained. Production of $Z^\prime$ bosons, followed by decays into $\tau$ lepton pairs have been also looked for, taking into account the fact that $\tau$'s can decay both leptonically and hadronically. A combined search of both leptonically and hadronically decaying $\tau$-pairs exclude $Z^\prime$ masses up to 2.42 TeV at $\sqrt{s}=13$ TeV with $36~{\rm fb}^{-1}$ luminosity \cite{Aaboud:2017sjh}. 

To insure the LHC constraints on the properties of $Z^\prime$ bosons are respected, we require our solutions to have a $Z^\prime$ mass heavier than 5 TeV as well as low $Z-Z^\prime$ mixings, $\mathcal{O}(10^{-3})$. Then we calculate the $Z^\prime$ production cross section. Using the decay table provided by the {\textsc SPheno} package and assuming the narrow-width approximation, we compare our predictions with the ATLAS and CMS limits on $Z^\prime$ bosons in the dilepton \cite{Aad:2019fac, Sirunyan:2021khd} and dijet \cite{Sirunyan:2018xlo, Sirunyan:2019vgj} modes in order to estimate the impact of supersymmetric decay channels in the non-universal $U(1)^\prime$ model. Following the methodology described above, we scan the parameter imposing constraints on SUSY particles, rare $B$-meson decays, $Z^\prime$ mass, both electron and muon magnetic moments within $1\sigma$ as indicated in Table \ref{tab:constraints}.

 Major contributions to the muon $g-2$ come from the neutralino-slepton and chargino-sneutrino loops. Thus, apart from the absolute constraints presented in Table \ref{tab:constraints}, LHC limits on the neutralino, chargino, slepton and sneutrino masses \cite{CMS:2021bra,Aad:2019byo,Aad:2019vnb,Aad:2019qnd} also impact the favored parameter region in this context. In this work we  considered scenarios where the LSP can be the lightest neutralino. The lightest sneutrino can also be the LSP and dark matter candidate. However, the sneutrino of course has to be dominantly singlet to be consistent with dark matter constraints. The lightest neutralino can be bino-dominated, wino-dominated or higgsino-dominated. Given the different LSP scenarios, the relevant mass limits on the electroweak sector particles will also be different depending on various factors, like the LSP mass and mixing, the LSP-NLSP mass gap and the available decay modes of the heavier particles (in particular the NLSP). 
\begin{itemize} 
\item For a bino-dominated LSP and wino-dominated NLSP, the most stringent constraint comes from their $WZ$ mediated decays. The experimental collaborations can exclude the wino-mass up to 650 GeV subjected to the bino mass. Similarly the LSP bino mass can be excluded up to at most 300 GeV  \cite{CMS:2021bra}. If the decay is $Wh$ mediated, the constraints are weaker. The bounds in that case on the wino and bino masses are around 250 GeV and 60 GeV respectively at 139 fb$^{-1}$ luminosity \cite{CMS:2021bra}.
\item If the SUSY spectra is such that one of the charged sleptons is the NLSP (along with a bino LSP), the heavier neutralino/chargino states may decay to the LSP via these sleptons. Depending on the NLSP slepton mass, the limits on the bino LSP and the heavier wino-like neutralino/chargino masses can be more severe. In such scenarios, the wino-like neutralino/chargino masses can be excluded up to 1450 GeV while the LSP bino mass can be excluded up to 900 GeV at 139 fb$^{-1}$ luminosity \cite{CMS:2021bra}. For $\tau$-enriched slepton mediated decays, the bounds are a bit weaker \cite{CMS:2021bra}.
\item Exclusion limits on the bino-mass are weaker for direct search of NLSP sleptons. If the NLSP slepton happens to be the lighter stau, the exclusion on the stau NLSP and bino LSP masses can go up to 400 GeV and 150 GeV respectively \cite{Aad:2019byo} at 139 fb$^{-1}$ luminosity. If the contributions from selectrons, smuons an staus are combined, the exclusion limit on their masses can go up to 700 GeV while the LSP can be ruled out up to 400 GeV \cite{Aad:2019vnb}. 
\item For higgsino LSP, the exclusion limits on the neutralino, chargino masses are much weaker. In this case, one can only rule out higgsino masses up to 250 GeV at most \cite{Aad:2019qnd}. 
\item The additional ${\tilde B}^\prime$  mixes with the singlino ${\tilde S}$, as seen from the neutralino mass matrix, Eq. \ref{eq:neutralino_mass}. Thus like later, it is constrained to be very heavy , due to the constaints on $Z^\prime$ mass ($M_{Z'}>5$ TeV) and not only can it never be the LSP, but its contribution is negligible.
\item For sneutrino LSP, the constraints are much weaker and no updated results from LHC are available for 13 TeV. One can put constraints on the neutralino, chargino and slepton states assuming sneutrino to be the LSP. Those constraints are comparable to the ones already discussed above. 
\end{itemize}

Given the fact that there are six neutralino states and additional sneutrino states, the present scenario is quite different from that of the MSSM. In the next section, we present the impact of these constraints on our favored parameter space (the one which satisfies both electron and muon $g-2$ experimental bounds). As it is computationally tedious to check the relevant constraints for every individual points in the scan,  the exclusion lines shown in the plots are often conservative.

\section{Analysis of lepton $g-2$ in non-universal $U(1)^\prime$ models}
\label{sec:LepG2}
In this section we present the numerical analysis for lepton anomalous magnetic moments, and investigate their dependence on the details of the spectrum.
 
In Fig. \ref{fig:DAMUvsDAEL} we plot, on the left,  points of the parameter space which satisfy both $\Delta a_\mu$ and $\Delta a_e$ to 1$ \sigma$ (blue circles), 2$ \sigma$ (green circles) and 3$ \sigma$ (olive green circles). In these plots we restrict the LSP to be the lightest neutralino and a viable dark matter candidate. 
The plot on the left
shows the impact of the new Fermilab data corresponding to $\Delta a_\mu$ on the pre-existing allowed regions. The new 1$ \sigma$, 2$ \sigma$ and 3$ \sigma$ allowed regions are shown by the light green, yellow and blue bands respectively while the allowed ranges from pre existing data are represented by the three red circles. These are combined likelihood regions that satisfy the $\Delta a_\mu$ and $\Delta a_e$ anomalies simultaneously within the three corresponding confidence levels. Evidently, the allowed regions have shrunk, which reduces some of the available parameter space. However, the plot indicates  that a large number of generated solutions lie in the desirable ranges. The plot on the right  of Fig. \ref{fig:DAMUvsDAEL} showcases the range of the non-holomorphic couplings, $T^\prime_\mu$ for smuons $T^\prime_e$ for selectrons, consistent with anomalous magnetic moments within 1-3$ \sigma$ of the measured values. There is clear indication that these couplings are required to have opposite sign. In fact, as we shall see later, both the magnitudes and signs of the non-holomorphic couplings are essential to yield different signs, and different magnitudes, for the electron and muon $g-2$. As expected from other analyses \cite{Demir:2005ti}, values of $T^\prime_\mu$ and  $T^\prime_e$ in the TeV range satisfy both mass and magnetic moment constraints for the leptons. 
As shall be seen in the next section, when we discuss the composition of the LSP and next-to-LSP (NLSP), the singlino and bino prime dominated neutralinos are always heavy. Thus, these neutralino states do not contribute significantly to the magnetic moments in this model. The contribution to $g-2$  instead depends heavily on the masses and couplings of the other neutralino, chargino states along with that of the sneutrinos, smuons and selectrons. The masses of selectrons and smuons are in turn affected by the choices of non-holomorphic couplings. Hence the non-universal structure of the model plays an essential role in obtaining $g-2$ results consistent with the experiment.
\begin{figure}
	\centering
	\includegraphics[scale=0.28]{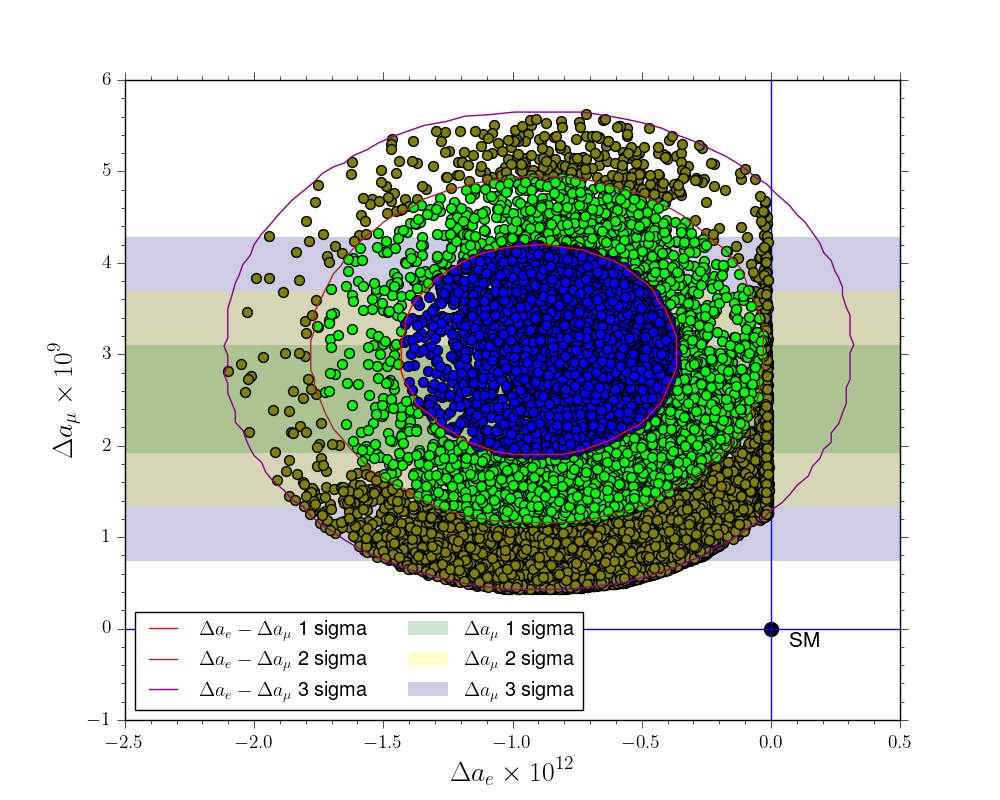}
	\includegraphics[scale=0.28]{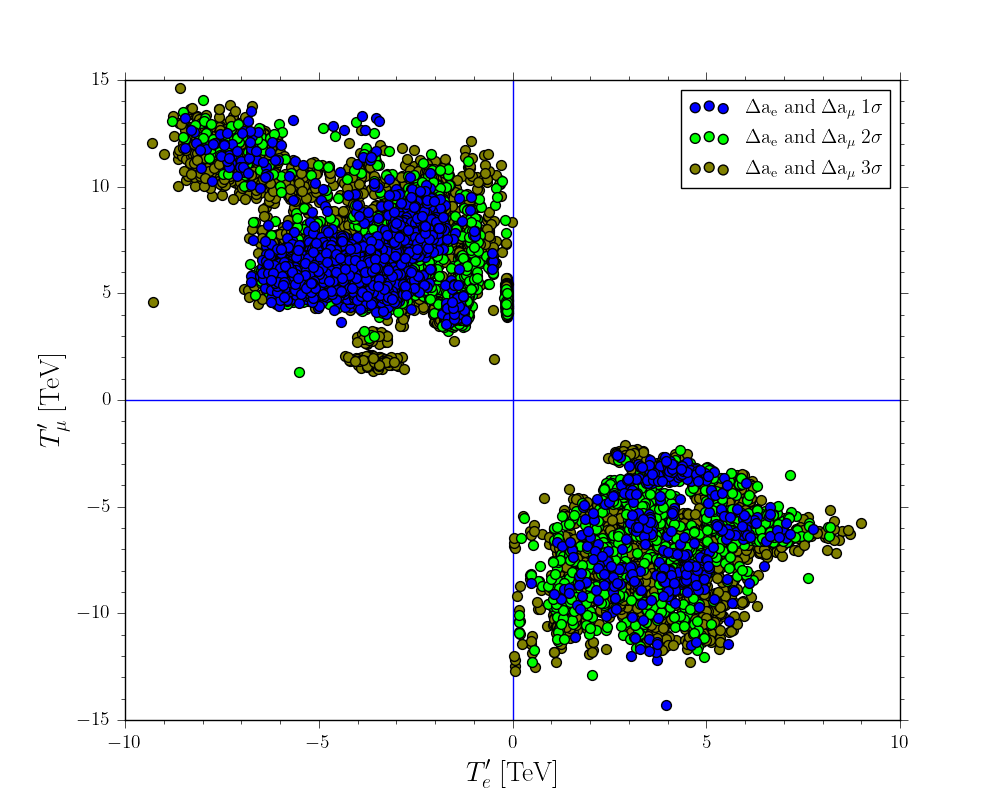} \\
	\caption{ Left: Parameter space points where $\Delta a_\mu$ and $\Delta a_e$ are within 1$ \sigma$ (blue circles), 2$ \sigma$ (green circles) and 3$ \sigma$ (olive green circles) of their experimental values; Right: anomalous magnetic moments dependence on the non-holomorphic couplings  $T^\prime_\mu$ and  $T^\prime_e$.}
	\label{fig:DAMUvsDAEL}
\end{figure}
 
In Fig. \ref{fig:U1Charges} we show correlations between the relevant non-universal $U(1)^\prime$ charges for left and right-handed selectrons and smuons such that the electron and muon magnetic moment contributions fall within the 3$\sigma$ range of the best fit value. 
The difference between required electron and muon $U(1)^\prime$ charges, besides the fact that they have different signs, indicate that the choices of $Q_{L_1}$ and $Q_{L_2}$  differ from each other and the same holds for $Q_{E^c_1}$ and $Q_{E^c_2}$. This result underlines the importance of non-universality in this scenario in order to explain the difference in lepton anomalous magnetic moments.
\begin{figure}
	\centering
	\includegraphics[scale=0.28]{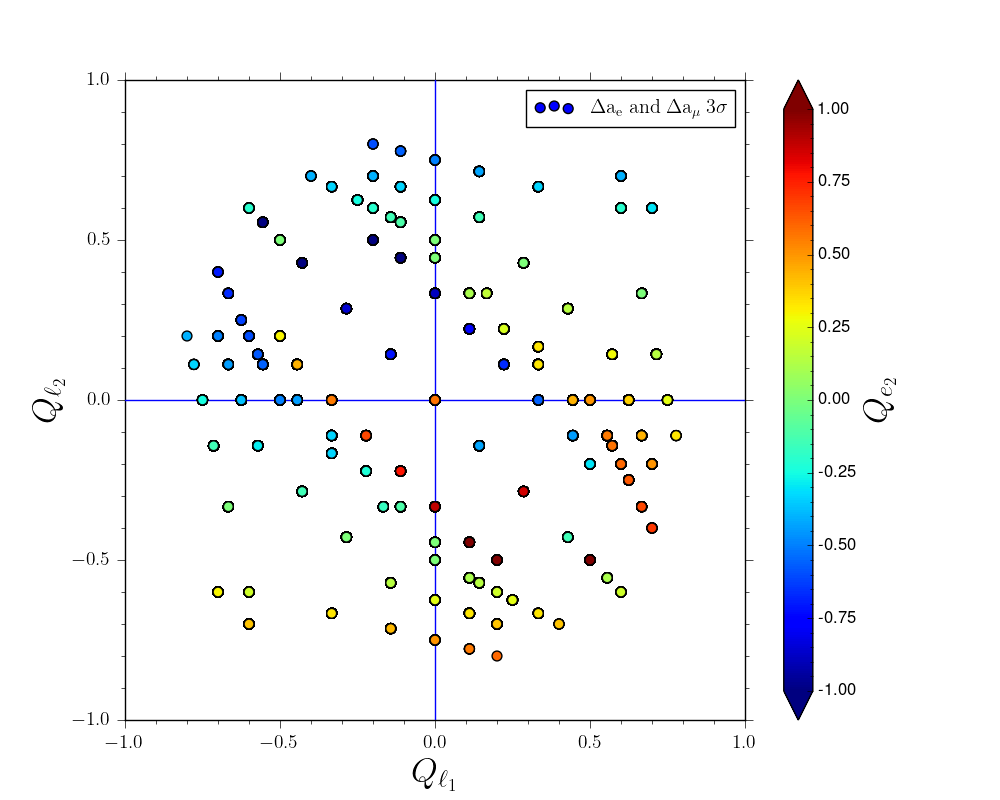}
	\includegraphics[scale=0.28]{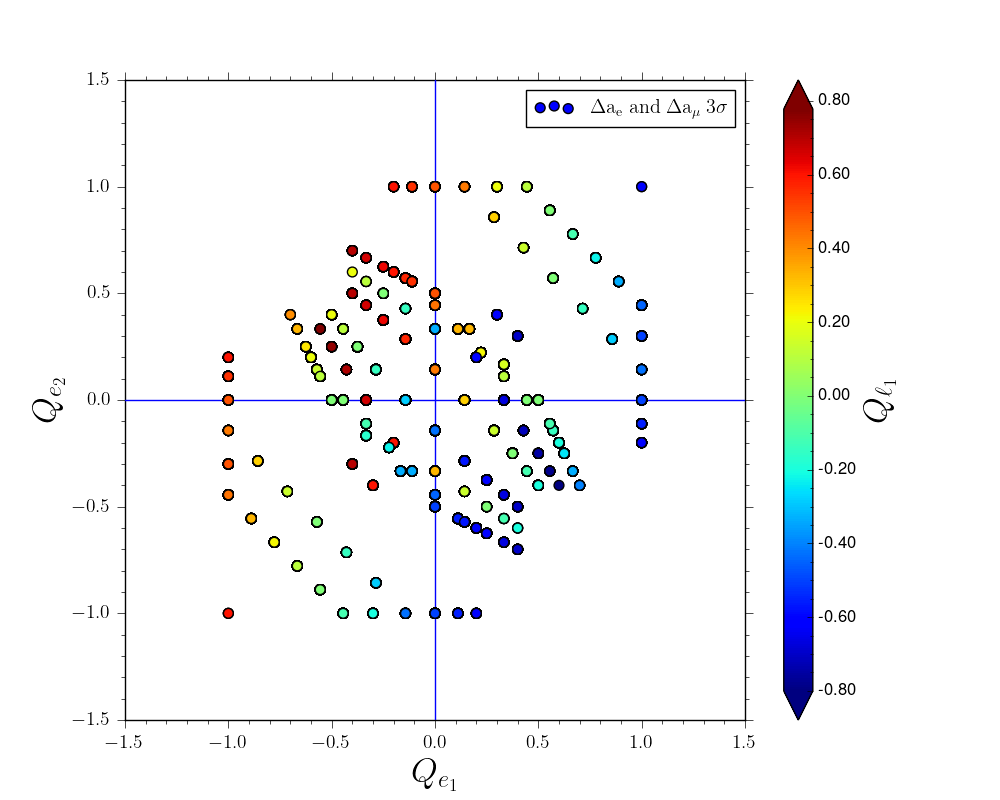}	\\	
	\caption{Correlations between electron and muon $U(1)^\prime$ charges, $Q_{L_1}$ and $Q_{L_2}$ (on left) and $Q_{E^c_1}$ and $Q_{E^c_2}$ (on right).}
	\label{fig:U1Charges}
\end{figure}

In Fig. \ref{fig:SoftMasses} we showcase the impact of the choices of gaugino and higgsino mass parameters on the calculation of $\Delta a_{e}$ and $\Delta a_{\mu}$. One can learn about the favoured ranges of these soft-SUSY mass parameters in order to obtain anomalous magnetic moments within their 1, 2 and 3$\sigma$ allowed ranges from the color-coded points corresponding to different LSP scenarios. The three plots represent scenarios with bino, wino and higgsino dominated\footnote{The LSP composition should be greater than 90\% bino, wino or higgsino for the respective cases.} LSP scenarios, respectively, from left to right. It is evident that a large range of solutions exist both below and above the TeV range with bino and wino dominated LSP scenarios with the higgsino dominated LSP scenario being slightly more restricted. Note also that there is a lack of apparent correlation between the choices of the soft-SUSY masses in the favoured parameter space and magnetic moment values within their 1, 2 and 3$\sigma$ allowed ranges. This is because of the fact that, along with the MSSM particle content, there are now three new sneutrinos and two new neutralino states, which give rise to a large number of possible diagrams contributing to the magnetic moment calculations. The contributions arising from processes with heavier neutralino and chargino states in the loop will understandably be smaller, but when all these contributions are added up, they can be significant. Especially, owing to the various complex mixing patterns one obtains in the neutralino sector while scanning, it is difficult to identify one or two processes that can explain the results. This is also the reason why a much larger range for these soft masses are consistent with $\Delta a_{e}$ and $\Delta a_{\mu}$ compared to situation in the MSSM.
We note that even if we choose the gaugino mass parameter $M_4$ to be light, the physical neutralino states involving ${\tilde B}^\prime$ are always heavy. This is due to mixing with the singlino, as in Eq. \ref{eq:neutralino_mass}, which is constrained to be heavy by the choice of $v_S$, which must yield a heavy $Z^\prime$ mass.
\begin{figure}
	\centering
	\includegraphics[scale=0.28]{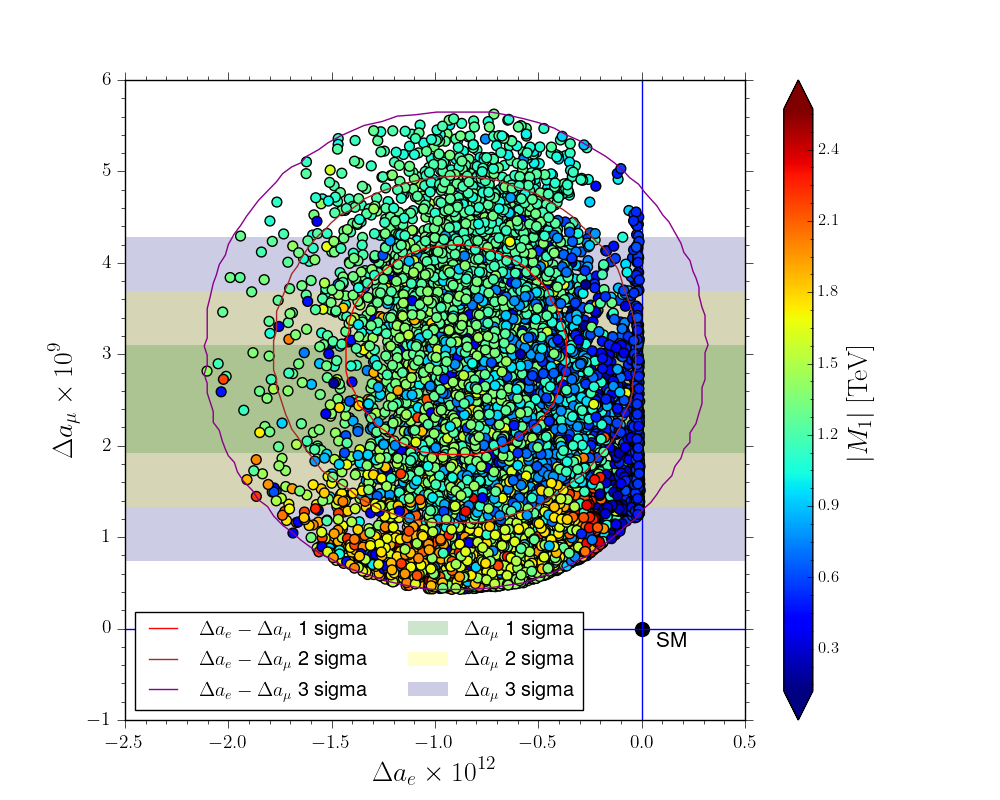} 
	\includegraphics[scale=0.28]{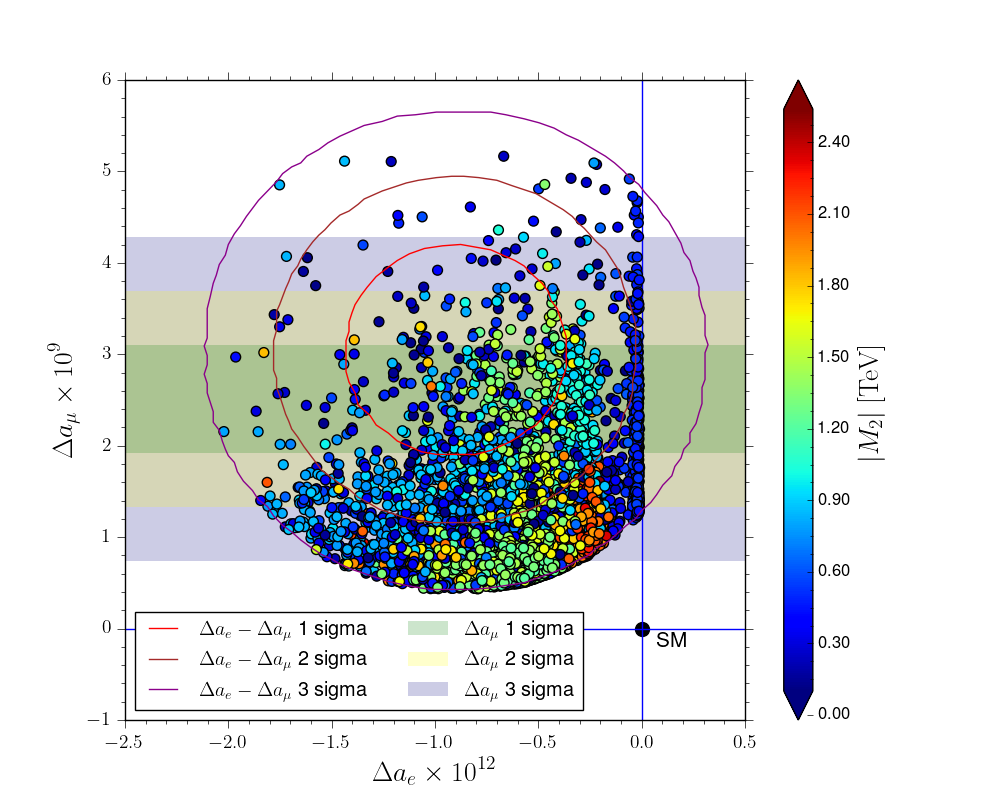} \\	
	\includegraphics[scale=0.28]{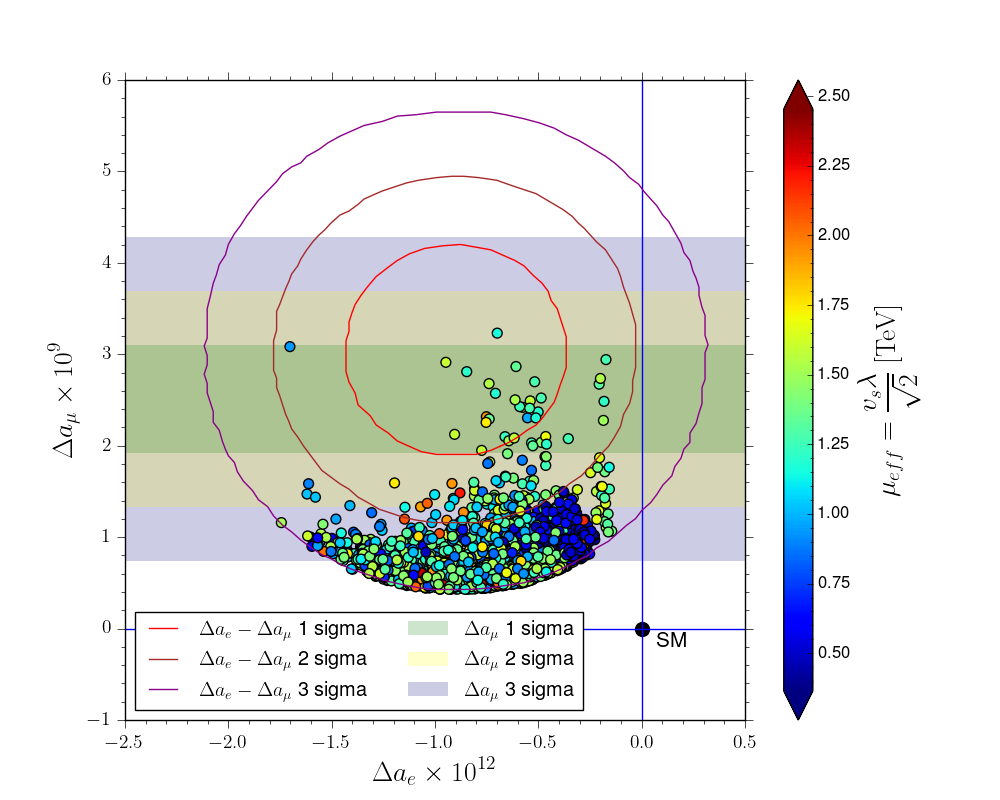} 	
		\caption{Gaugino/higgsino mass parameters dependence on $\Delta a_\mu$ and $\Delta a_e$. We show solutions which pass all constraints, with masses indicated in the right-handed legends. Top left: bino-dominated solutions, Top right: wino-dominated solutions, Bottom:  higgsino-dominated solutions
		.}
	\label{fig:SoftMasses}
\end{figure}

 However, a significant contribution always arise from the lightest neutralino-lightest slepton and lightest chargino-lightest sneutrino loops. So, it is worth investigating the ranges of the physical masses of these lightest states. 
In Fig. \ref{fig:Spectrum} we present the correlations between the lightest neutralino-lightest slepton masses (left) and lightest chargino-lightest sneutrino masses (right). Evidently, the most favored region for the lightest neutralino state is $\sim$ 0.2 - 1.3 TeV while that for the corresponding lightest charged slepton it is $\sim$ 0.5 - 3.0 TeV. 
 However, as the plot indicates there are some parts of the parameter space where we can get a consistent solution with lighter charged sleptons as well. In our scan the soft slepton masses for the three families are assigned at SUSY scale independent of each other \cite{Araz:2017wbp}. Therefore, the lightest slepton can be the smuon or selectron, rather than stau, and  this explains why we can fulfill lepton $g-2$ constraints to 1$\sigma$ easily. Understandably, the acceptable chargino mass range is wider, $\sim$ 0.2 - 3.0 TeV while that for the sneutrino is $\sim$ 0.5 - 3.6 TeV. 
\begin{figure}
	\centering
	\includegraphics[scale=0.28]{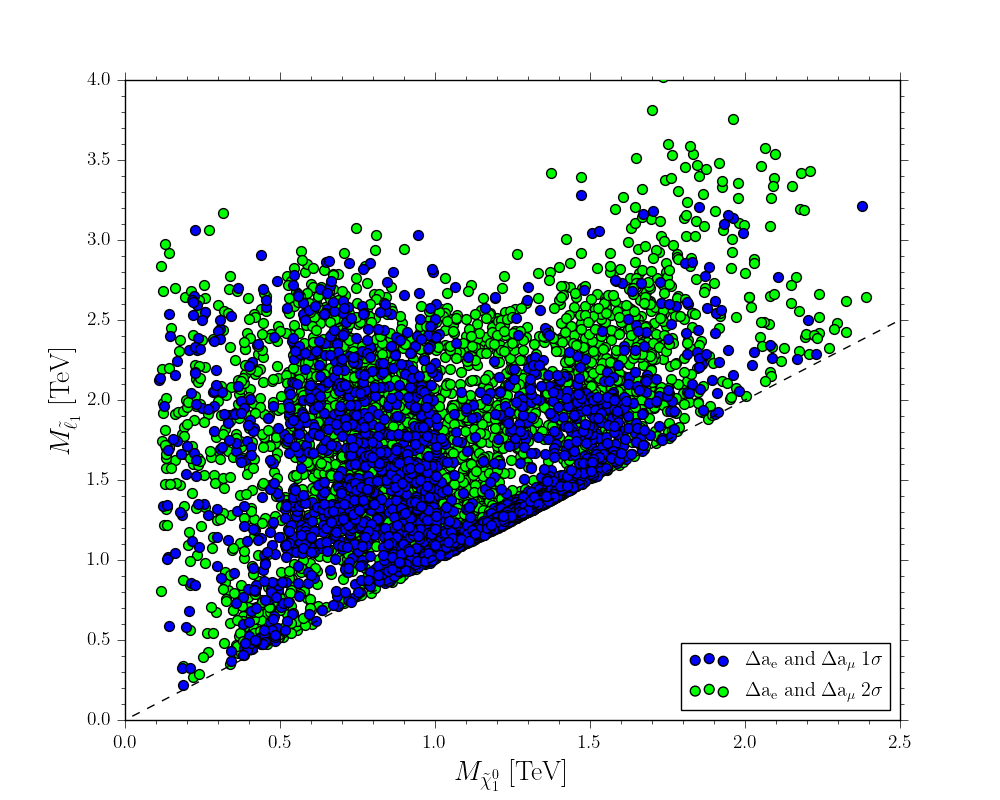}		
	\includegraphics[scale=0.28]{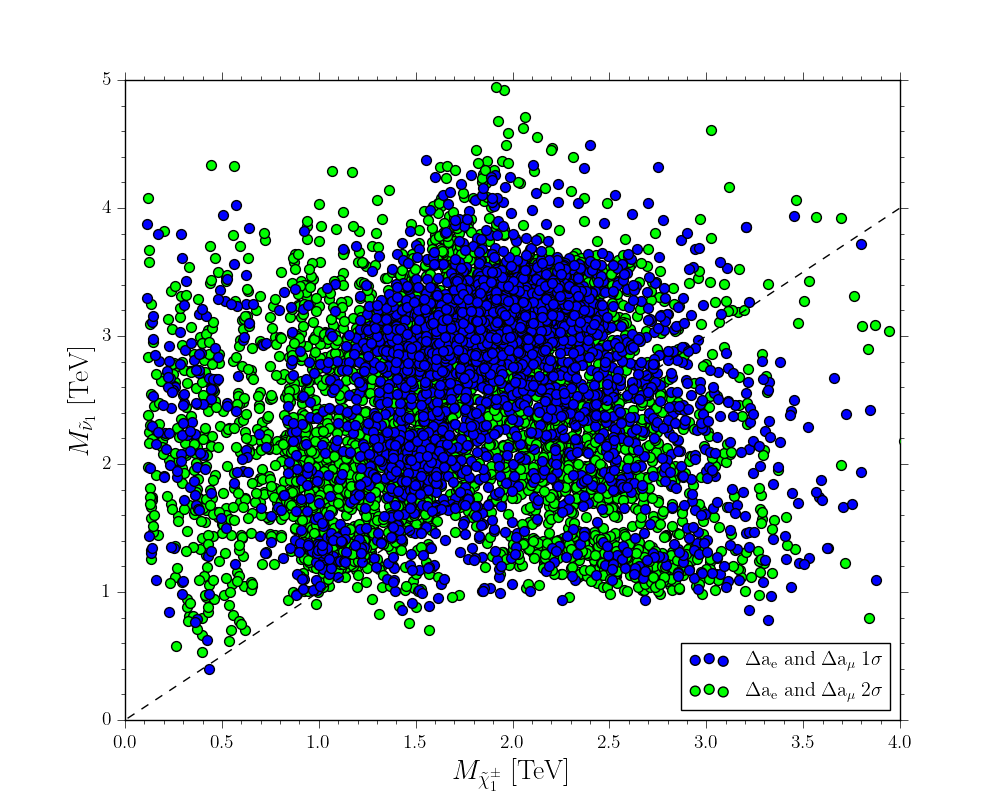}	
	\caption{Correlation between the choices of lightest neutralino-lightest slepton masses {left} and lightest chargino-lightest sneutrino masses {right} in order to obtain $\Delta a_\mu$ and $\Delta a_e$ within 1 $ \sigma$ (blue circles), 2$ \sigma$ (green circles) of their experimental values. All possible LSP scenarios are included.}
	\label{fig:Spectrum}
\end{figure}
In the next section, we study the impact of LHC direct SUSY search constraints and the dark matter constraints on the favored parameter space discussed in the previous section.

\subsection{LHC Constraints}
\label{sec:Chi1LSP_LHC}

The impact of the LHC constraints on the electroweak sector particle masses varies depending on the nature of the LSP, NLSP and the composition of the relevant states as discussed in detail in section \ref{sec:WpZpmass}. As a case study, in Fig. \ref{fig:lhc_const1} we show the impact of the relevant constraints on the LSP neutralino and NLSP chargino mass plane. Here the LSP is bino dominated while the NLSP is wino dominated. Hence the lighter chargino mass is nearly degenerate with the second lightest neutralino mass. This scenario presents the most restricted parameter space thus far. However, as can be seen from the figure, although it has a sizeable  impact on the available parameter space, in our scenario, a large portion of it remain unexplored. The exclusion lines are taken from public results of CMS experimental collaboration \cite{CMS:2021bra}. As discussed before, ATLAS collaboration also provides similar constraints. Note that, some of the points ruled out by the exclusion lines may not be excluded after all since while scanning the parameter space we did not assume the simplified model scenarios considered by the experimental collaborations to present their results. Therefore, Fig. \ref{fig:lhc_const1} represents a conservative estimation and if anything, the impact of the constraints in the present context will be weaker.
\begin{figure}
	\centering
	\includegraphics[scale=0.35]{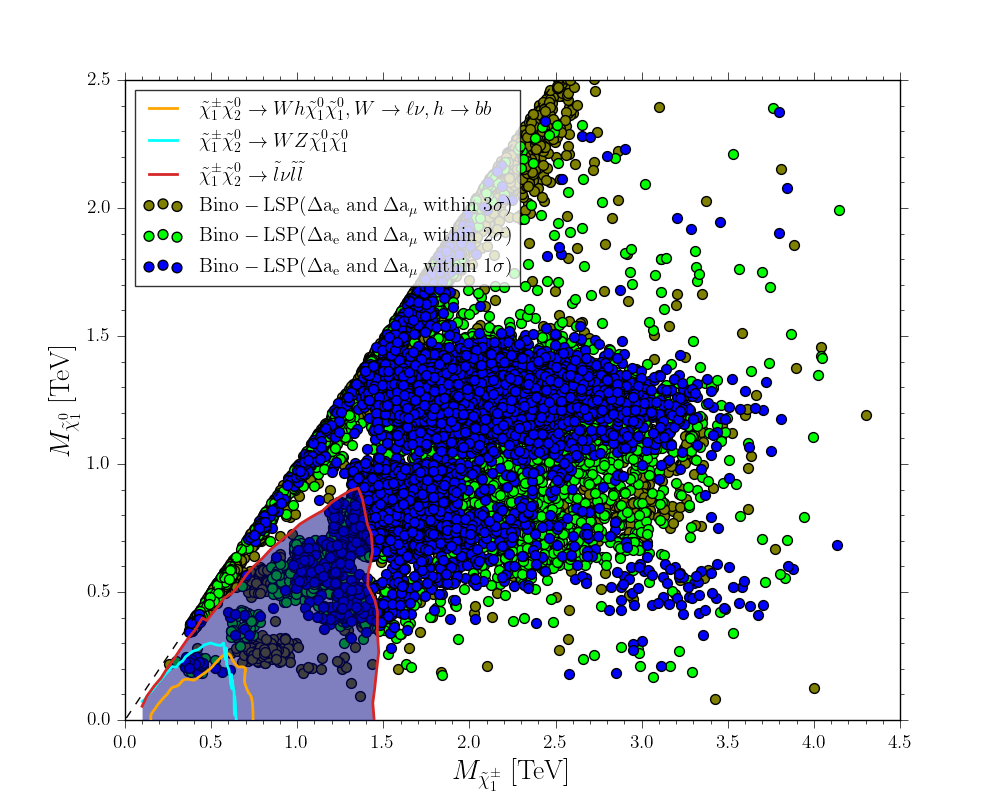}		
	\caption{The favored parameter space with bino dominated LSP along with the LHC constraints derived from direct search.}
	\label{fig:lhc_const1}
\end{figure}

\subsection{Dark Matter Constraints}
\label{sec:Chi1LSP_DM}
For the calculation of lepton magnetic moments,  $g-2$ and dark matter constraints were not correlated. In Fig. \ref{fig:Chi1LSP} we  divide the dark matter content into three cases, depending on whether it is higgsino, bino, or wino dominated. In no cases do we get ${\tilde B}^\prime$ domination, even for small bino prime masses. The reason, as previously discussed, can be seen from analysing the neutralino mass matrix, Eq.  \ref{eq:neutralino_mass}. While the bino mass mixes with the (relatively lighter) higgsinos ${\widetilde H}_u$ and  ${\widetilde H}_d$, the bino prime mass includes, in addition,  mixing with the heavier singlino  ${\widetilde S}$. In the left panel of Fig. \ref{fig:Chi1LSP}, we verify the relic density for bino,  higgsino or wino dominated LSP.  We note that like in MSSM, the relic density is dominated at lower masses by the bino LSP,  while higgsino LSP saturates the Planck value \cite{2020} between 900 - 1400 GeV. Wino LSP points are required to have masses in excess of $1.5$ TeV to yield correct relic density.  We further checked if we can obtain a good benchmark, and found higgsino-dominated LSP solutions which yield lepton anomalous magnetic moments for both  muon and electron within 1$\sigma$ of the experimental value. Wino-dominated dark matter candidates which obey relic constraints are heavier and may cross the line between 1.5 - 2.3 TeV, but below this value they are underabundant.
Many viable solutions are bino-dominated, and while most yield overabundant relic densities,  some satisfy the constraints,  mostly those which are mixtures of binos and higgsinos.  We also explore limits from direct detection. We show, in Fig. \ref{fig:Chi1LSP} right panel, the limits from cross sections with protons. Of the solutions that satisfy relic constraints only a small portion lies in the excluded region and the resulting parameter points are consistent with the existing direct detection constraints. Neutron direct detection cross sections are exactly the same, so we do not show them separately. We note that bino-like solutions are most promising, as they fall significantly below the Xenon1T constraints \cite{Aprile_2019, Aprile_2017} and even below constraints from Darwin \cite{Aalbers_2016}. Wino-like solutions can also satisfy most direct detection constraints, while most higgsino-like LSP are just below but close to the Xenon1T limits and will likely be ruled out by more precise experiments in the near future.

\begin{figure}
	\centering
	\includegraphics[scale=0.28]{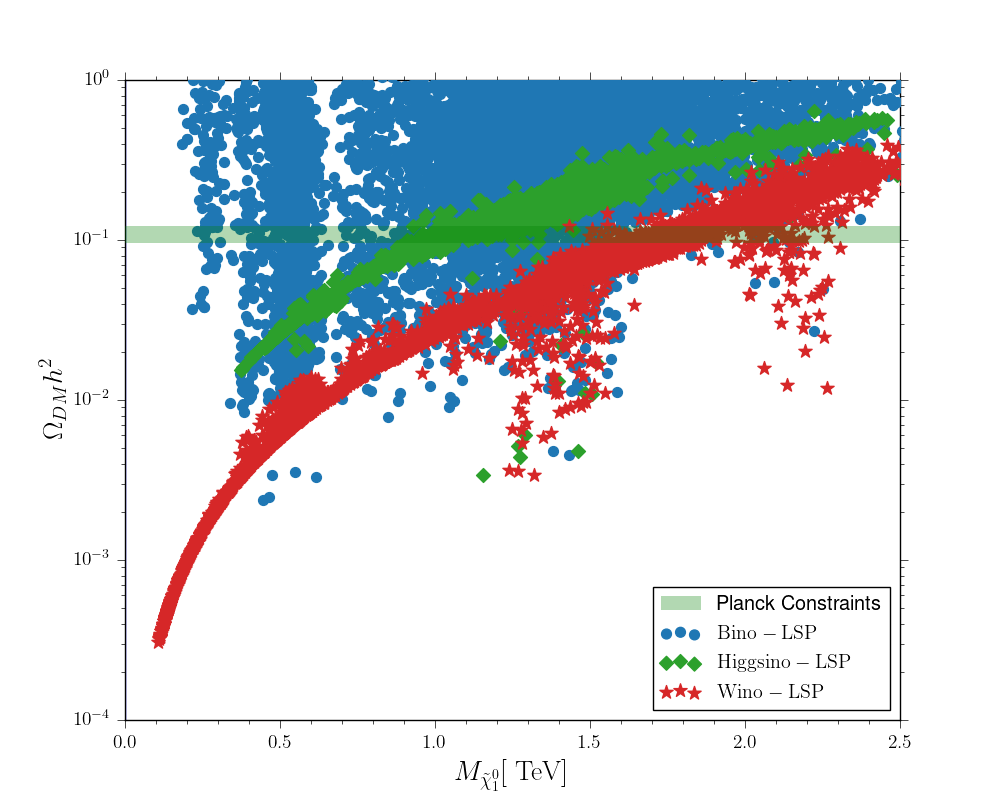}
	\includegraphics[scale=0.28]{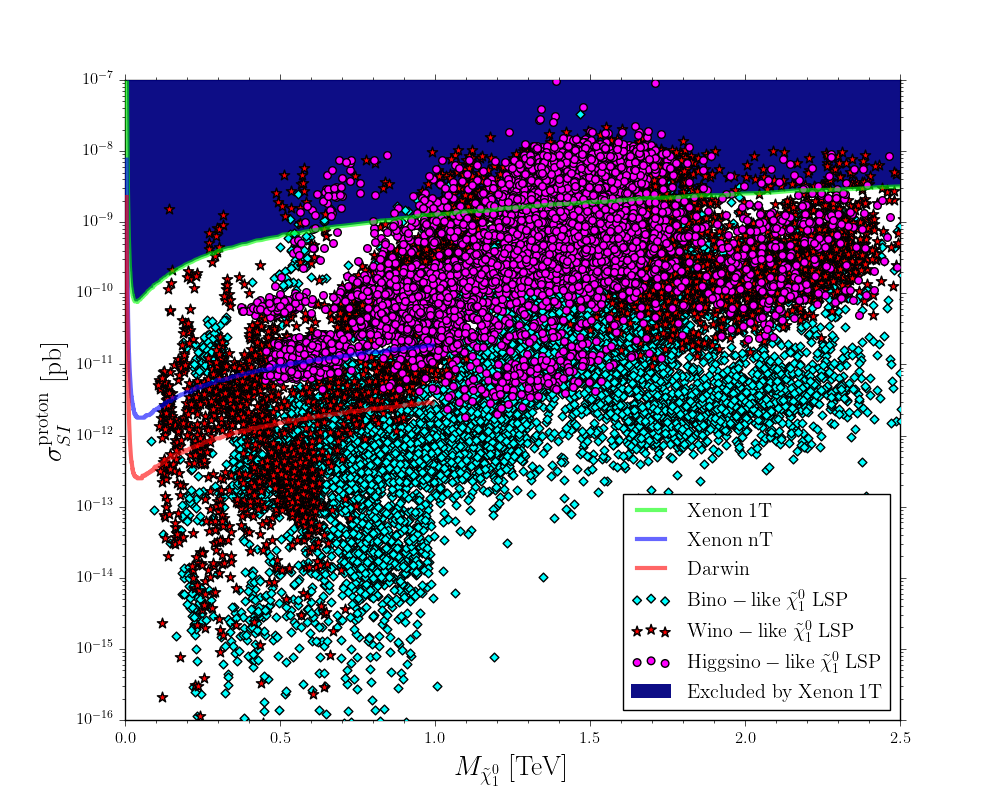}	
	\caption{Neutralino dark matter. Left panel: relic density plot for bino, higgsino or wino dominated neutralino LSP, compared to the measured values by Planck \cite{2020}. Right panel: direct detection limits for neutralino LSP,  depending on its composition, compared to the Xenon1T  \cite{Aprile_2019, Aprile_2017} and Darwin \cite{Aalbers_2016} constraints. }
	\label{fig:Chi1LSP}
\end{figure}

In this model, the LSP can also be the sneutrino. A more detailed scan around the resonance regions yields points with right-handed sneutrino LSP satisfying relic density requirement, However in these cases, the sneutrino DM sector behaves exactly like MSSM + RH neutrino models, and thus, in what follows, we analyse only cases where the neutralino is the LSP.

\section{Model characterization at LHC}
\label{sec:Collider}
In this section, we investigate the observability of a non-universal UMSSM scenarios with heavy $Z^\prime$ masses at LHC, in particular when the high-luminosity LHC run is considered.  We are interested in finding ways to differentiate this model from other $U(1)^\prime$ models, in particular, from $U(1)^\prime$ models with universal couplings. To determine the signals to be searched for, we focus on a set of promising benchmarks obtained from our scan results for which all constraints are satisfied. In order to evaluate the fiducial cross sections associated with Drell-Yann signals, we export the non-universal UMSSM to the UFO format \cite{Degrande:2011ua} and make use of the \mg~ framework (version 3.0.3)  \cite{Alwall:2014hca}  to simulate hard-scattering LHC collisions. These  events, obtained by convoluting the hard-scattering matrix elements with the LO set of NNPDF 3.1 parton densities \cite{Ball:2017nwa} are analysed within the \ma~ framework (version 1.8.58).

To highlight the model characteristics, we focus on three optimistic signal benchmarks, {\bf BM I}, {\bf BM II} and {\bf BM III}, that are currently not excluded by data and with different $U(1)^\prime$ properties. All three scenarios exhibit a $Z^\prime$ boson with a mass of 5.5 TeV and $\sigma(pp \to Z^\prime) \times BR (Z^\prime \to \ell \ell)$ as large as possible, in order to maximize the total cross section. All benchmarks are consistent with all constraints, including relic density, and satisfy the bounds on the $g-2$ factor of the electron and muon at $1 \sigma$. The LSP neutralino is bino, wino and higgsino  for  {\bf BM I}, {\bf BM II} and {\bf BM III}, respectively. 

In Table \ref{tab:benchmark_free} we present a complete list of the three benchmark model parameters, including gaugino masses, non-holomorphic couplings, $U(1)^\prime$ gauge coupling, and the non-universal charges for all the particles in the model. We note that the latter are very close among the three benchmarks, indicating the strict constraints imposed by the anomalous magnetic moments. 
In Table \ref{tab:benchmarks_mass} we give the values of the masses of the particles in the model for the three scenarios. Note that here dark matter and $g-2$ constraints are satisfied for relatively heavy superpartner masses, and thus the spectrum is consistent with strong constraints obtained from the Run-1 and Run-2 at the LHC.
Dark matter and anomalous magnetic moment values for the three benchmarks are given in Table \ref{tab:BenchmarkRelic}, indicating that relic density is kept within 5$\sigma$ of its measured value and the direct detection cross section is below its upper bound at Xenon1T. In Table \ref{tab:lhc} we show properties of the $Z^\prime$ gauge boson in our model, including production cross sections at 13, 14, 27 and 100 TeV, and branching ratios into chargino and neutralino pairs, jets, leptons and neutrinos. The surprising equality between the branching ratios of $Z^\prime$ to electron and muon pairs is due to the choice of $U(1)^\prime$ charges in Table \ref{tab:benchmark_free}, where right-handed electron charges match left-handed muon ones, and right-handed muon charges are the same as left-handed electron charges.  Therefore, the electron and muon for each benchmark have same vector and axial-vector $ Z^\prime  $ couplings that impact the branching ratio of $ Z^\prime  $. The only non-universality is then present in the $Z^\prime \to \tau^+ \tau^-$ decay, which differs from the branching ratio $Z^\prime \to \mu^+ \mu^-, e^+e^-$.

\begin{table}[t]
	\renewcommand{\arraystretch}{1.3}\setlength\tabcolsep{6pt}
	\begin{center}
		\begin{tabular}{c|c c c c c c c }
			\small
			[GeV] &$M_1$& $M_2$ & $M_3$ & $M_4$ & $v_S$ & $g^\prime_{\rm SUSY}$ \\
			\hline\hline
			{\bf BM I}   & 453   & -1088  & 31858 & 300  & 12573 & 0.462  \\
			{\bf BM II}  & 1614  & 1509   & 7568   & 1849 & 14964 & 0.472  \\					
			{\bf BM III}  & -1967 & 2139  & 8109  & -3087 & 15172 & 0.472  \\			
		\end{tabular}\\[.2cm]
		\begin{tabular}{c|c c c c c c c c c}
			\small
			& $\tan \beta$   & $\lambda$  & $T_\lambda$
			& $T_e^\prime$ & $T_\mu^\prime$
			& $A_\tau $ & $Y_v^{ij}$ \\
			\hline\hline
			{\bf BM I}   & 22.8 & 3.01$\times 10^{-1}$ &-794   &  -4199  & 6249 & 178  & $10^{-8}$ \\
			{\bf BM II}  & 40.6 & 2.43$\times 10^{-1}$ & 1167  & -2106  & 9454 &-399  & $10^{-8}$ \\					
			{\bf BM III}  & 33.0 & 9.07 $\times 10^{-2}$ & 13122 & 4382 & -7300 & -41.15  & $10^{-8}$ \\			
		\end{tabular}\\[.2cm]	
		\begin{tabular}{c|c c c c c c c c c}
			\small
			& $Q_{\ell_1}$   & $Q_{\ell_2}$ & $Q_{\ell_3}$
			& $Q_{e_1}$ & $Q_{e_2}$	& $Q_{e_3}$ & $Q_{\nu_1}$ & $Q_{\nu_2}$ & $Q_{\nu_3}$ \\
			\hline\hline
			{\bf BM I}   & -0.11 & 0.55  & -0.44 & 0.55  & -0.11 & -0.11 & -0.33 & -1.0 & 0.  \\
			{\bf BM II}  &  0.60 & -0.20 & -0.40 & -0.20 & 0.60 & -0.10  & -1.0 & -0.20 & 0. \\					
			{\bf BM III}  &  -0.20 &  0.60 & 0.40 & 0.60 & -0.20 & -0.10 & -0.20 & -1.00 & 0. \\			
		\end{tabular}\\[.2cm]	
		\begin{tabular}{c|c c c c c c}
	\small
	& $Q_{q}$   & $Q_{u}$ & $Q_{d}$
	& $Q_{H_u}$ & $Q_{H_d}$	& $Q_{S}$ \\
	\hline\hline
	{\bf BM I}   & -0.11 & -0.33 & -0.44 & 0.44 & 0.55 & -1.0   \\
	{\bf BM II}  &-0.10 & -0.30 & -0.40 & 0.40 & 0.50 & -0.90  \\	
	{\bf BM III}  & -0.10 & -0.30 & -0.40   & 0.40  & 0.50 & -0.90  \\				
\end{tabular}\\[.2cm]	
		\caption{Set values for the free Non-UMSSM parameters defining our 
		benchmark scenarios {\bf BM I},  {\bf BM II} and {\bf BM III}. All masses are given in GeV.}
		\label{tab:benchmark_free}
	\end{center}
\end{table}

\begin{table}[t]
	\renewcommand{\arraystretch}{1.3}\setlength\tabcolsep{6pt}
	\begin{center}
		\begin{tabular}{c|c c c c c c c}
			[GeV] & $M_{Z'}$ & $M_{H_1^0}$   & $M_{H_2^0}$  & $M_{H_3^0}$ &  $M_{A_1^0}$   & $M_{H_1^\pm}$ \\
			\hline\hline
			{\bf BM I}   & 5500 & 126  & 5734  & 15260 & 15260  & 15266 &   \\
			{\bf BM II}  & 5500 & 123  & 6340  & 22405 & 22405 & 22418 &    \\				
			{\bf BM III}  & 5500 & 122 & 6441  & 21073  & 21073 &  21075  &    \\								
		\end{tabular}\\[.2cm]
		\begin{tabular}{c| c c  c c c c c c c c}
			[GeV]  &  $ \tilde{\chi}_1^0 $ Comp.  & $M_{\tilde{\chi}_1^0}$
			& $M_{\tilde{\chi}_2^0}$  & $M_{\tilde{\chi}_3^0}$ & $M_{\tilde{\chi}_4^0}$
			& $M_{\tilde{\chi}_5^0}$  & $M_{\tilde{\chi}_6^0}$  & $M_{\tilde{\chi}_1^\pm}$ & $M_{\tilde{\chi}_2^\pm}$ & $M_{\tilde{g}}$\\
			\hline\hline
			{\bf BM I}   & Bino-like        & 442  & 1192 & 2747  & 2748  & 5581 & 5875 & 1192 & 2749  & 32043  \\
			{\bf BM II}  & Wino-like      & 1595 & 1601 & 2589  & 2593 & 5478 & 7317  & 1595 & 2592 & 7820   \\					
			{\bf BM III}  & Higgsino-like & 992 & 995  & 1949 & 2240 & 6275 & 6583  & 993   & 2240 & 8383 \\									
		\end{tabular}
		\begin{tabular}{c|c c c c c c c c c c c c}
			[GeV]  & $M_{\tilde{d}_1}$ & $M_{\tilde{d}_2}$ & $M_{\tilde{d}_3}$ & $M_{\tilde{d}_4}$
			& $M_{\tilde{d}_5}$  & $M_{\tilde{d}_6}$ & $M_{\tilde{u}_1}$ & $M_{\tilde{u}_2}$ & $M_{\tilde{u}_3}$ & $M_{\tilde{u}_4}$ & $M_{\tilde{u}_5}$ & $M_{\tilde{u}_6}$\\
			\hline\hline
			{\bf BM I}   & 23730 & 24305 & 24789 & 24789 & 25870  & 25870 & 21125 & 23732 & 25127 & 25127 & 25870 & 25870  \\
			{\bf BM II}  & 4740   & 6192    & 7537  & 7538    & 9300    & 9301   & 4727  & 5140   & 6771    & 6771   & 7536   & 7537    \\			
			{\bf BM III}  & 5926  & 7348 & 7879 & 7880  & 9247  & 9247 & 5439 & 5936 & 7155 & 7155  & 7879 & 7880 \\									
		\end{tabular}
		\begin{tabular}{c|c  c c c c c c c c c c c}
			[GeV]  & $M_{\tilde{\ell}_1}$ & $M_{\tilde{\ell}_2}$ & $M_{\tilde{\ell}_3}$ & $M_{\tilde{\ell}_4}$
			& $M_{\tilde{\ell}_5}$  & $M_{\tilde{\ell}_6}$ & $M_{\tilde{\nu}_1}$ & $M_{\tilde{\nu}_2}$ & $M_{\tilde{\nu}_3}$ & $M_{\tilde{\nu}_4}$ & $M_{\tilde{\nu}_5}$ & $M_{\tilde{\nu}_6}$ \\
			\hline\hline
			{\bf BM I}  & 479  & 2961 & 3206  & 3796 & 4169 & 5831 & 2986 & 3216  & 3785 & 4777 & 5366 & 5831  \\
			{\bf BM II} & 1623 &3561  & 4012  & 4514 & 4790 & 5374 & 3562 & 4012  & 4159 & 4587 & 5364 & 6437  \\					
			{\bf BM III} & 1598 & 2435 & 4143 & 4988 & 5314 & 5597 & 1013 & 1619 &3909&5308 & 5596 & 7292  \\									
		\end{tabular}				
		\caption{Particle spectrum of {\bf BM I},  {\bf BM II} and {\bf BM III}: bosons (top), fermions (middle), squarks and sleptons (bottom). All masses are given in GeV.}
		\label{tab:benchmarks_mass}
	\end{center}
\end{table}

\begin{table}
	\renewcommand{\arraystretch}{1.3}\setlength\tabcolsep{6pt}
	\begin{center}
		\small		
		\begin{tabular}{c|c c c c | c c}
			& $\Omega_{\rm DM} h^2$
			& $\sigma_{\rm SI}^{\rm proton}$ [pb]
			& $\sigma_{\rm SI}^{\rm neutron}$ [pb]
			& $\langle\sigma v\rangle$ [cm$^3$s$^{-1}$]
			& $\Delta a_e \times 10^{12} $
		 	& $\Delta a_{\mu} \times 10^{10} $	\\
			\hline\hline
			{\bf BM I}  & 0.091 & 2.93 $\times 10^{-13}$ & 2.97 $\times 10^{-13}$
			& 3.95$\times 10^{-29}$  & -1.40 (within 1$\sigma$) & 26.97 (within 1$\sigma$) \\
			{\bf BM II}  & 0.102 & 1.12 $\times 10^{-10}$ & 1.13 $\times 10^{-10}$ & 3.08 $\times 10^{-26}$ & -0.39 (within 1$\sigma$) & 30.55 (within 1$\sigma$) \\				
			{\bf BM III}  & 0.114 & 6.32 $\times 10^{-11}$ & 6.40 $\times 10^{-11}$ &  1.07$\times 10^{-26}$ & -0.55  (within 1$\sigma$) & 23.05  (within 1$\sigma$) \\									
		\end{tabular}
		\caption{Predictions  for the {\bf BM I},  {\bf BM II} and {\bf BM III}
			scenarios, of the observables discussed in our dark matter and lepton $g-2$ analysis.}
		\label{tab:BenchmarkRelic}
	\end{center}
\end{table}
\begin{table}
	\renewcommand{\arraystretch}{1.3}\setlength\tabcolsep{6pt}
	\begin{center}
		\small
		\begin{tabular}{c|cccc|c c c }
			&$\sigma (pp \to Z')$ [fb]	 &&& & BR($Z'\to \tilde{\chi}_1^\pm \tilde{\chi}_1^\mp$) & BR($Z'\to j j$)  & BR($Z'\to \ell \ell$)  \\
			\hline\hline
			& 13 TeV  & 14 TeV & 27 TeV & 100 TeV &&&\\ \hline
			{\bf BM I}  & 0.1795 & 0.2945 & 9.398 & 324.8  & 3.01 $\times 10^{-7}$ & 0.46 & 0.11 \\
			{\bf BM II}  & 0.1515 & 0.2493 & 7.933 & 274.7 & 8.07 $\times 10^{-7}$ & 0.41  & 0.15 \\										
			{\bf BM III} & 0.1520  & 0.2494  & 7.977  & 275.9 & 5.56 $\times 10^{-2}$ & 0.36 & 0.13  \\							
		\end{tabular}\\[.2cm]	
		\centering
		\begin{tabular}{c|c c c c}
			& BR($Z'\to e e$)
			& BR($Z'\to \mu \mu$)
			& BR($Z'\to \tau \tau$)
			& BR($Z'\to \nu_i \bar{\nu}_i$) \\
			\hline\hline
			{\bf BM I}  & 5.68 $\times 10^{-2}$  & 5.68 $\times 10^{-2}$  & 3.71 $\times 10^{-2}$  & 0.28   \\
			{\bf BM II} & 7.62 $\times 10^{-2}$ & 7.62 $\times 10^{-2}$  & 3.24 $\times 10^{-2}$  &  0.30 \\				
			{\bf BM III} & 6.70 $\times 10^{-2}$  & 6.70 $\times 10^{-2}$ &  2.84 $\times 10^{-2}$ & 0.27  \\									
		\end{tabular}
		\caption{$Z^\prime$ production cross section at $\sqrt{s}=13, 14, 27 $ and $100$ TeV and branching ratios  for the {\bf BM I},  {\bf BM II} and {\bf BM III} scenarios,  relevant for the associated LHC phenomenology.}
		\label{tab:lhc}
	\end{center}
\end{table}

We further explore the model consequences in Fig. \ref{fig:Zprime_limits}, where in the top panel we show properties of the $Z^\prime$ boson in this model. On the left, we plot the production cross section for $Z^\prime$ times the dilepton ($\ell=e, \mu$) branching ratio. For masses $M_{Z^\prime} \ge 5.5 $ TeV, mass bounds are satisfied, even for dilepton branching ratios of ${\cal O}(25 \%)$. Also values of the $U(1)^\prime$ coupling constant $g^\prime$, evaluated at SUSY scale, can be as large as $\sim 0.5$ (right plot).

On the top left of Fig. \ref{fig:Zprime_pheno} we show the dilepton invariant mass $M_{ll}$ with only basic cuts ($\lvert \eta_\ell  \rvert<$ 2.5, 5 TeV $< M_{\ell \ell} <$ 6 TeV) and compare  the results of our model with predictions from $E6$ motivated $U(1)^\prime_\psi$ and $U(1)^\prime_\eta$. For all models, the  branching ratios and production cross section of $Z^\prime$ are very similar. Thus, our model  cannot be distinguishable from other scenarios. Due to the limited SM number of events, the significance is low for both our non-universal $U(1)^\prime$ model and the other models. For characterization of signals, we try other methods for detection \cite{Araz:2021dga}. On the top right panel of Fig. \ref{fig:Zprime_pheno}, we show  the forward-backward asymmetry for  all the $U(1)^\prime$  models from the left-hand plot. The forward-backward asymmetry in $pp \to (Z / Z^\prime / \gamma) \to \ell^+ \ell^-$, ($\ell=e, \mu$) is
\begin{equation}
A_{\rm FB}=\frac{ d \sigma/dM(\ell^+ \ell^-)|\eta(\ell^-)>0|- d \sigma/dM(\ell^+ \ell^-)|\eta(\ell^-)<0|} {d \sigma/dM(\ell^+ \ell^-)|\eta(\ell^-)>0|+ d \sigma/dM(\ell^+ \ell^-)|\eta(\ell^-)<0|}
\end{equation}
where $M(\ell^+ \ell^-)$ is the neutral current (NC) Drell-Yan (DY) lepton pair invariant mass and $\eta(\ell^-)$ is charged lepton pseudorapidity while the identification of the forward (F) and backward (B) hemispheres via the restrictions $\eta(\ell^-)>0$ and $\eta(\ell^-)<0$.

The dashed lines are for  $U(1)^\prime_\psi$ and $U(1)^\prime_\eta$, the solid lines are for the three benchmarks in our model. We observe that different models have unique curves, and in our model, the asymmetry does not change from one BM to another: we have the same $M_{Z^\prime}$; $g^\prime$, $Q_u$, and $Q_d$, which are important for the process, are very close. This indicates that the model is robust in asymmetry predictions. We calculate the Kullback-Leibler divergence, $D_{\rm KL}$, as measure of how the probability distribution of our scenarios differs from the probability distributions of $U(1)^\prime_\psi$ and $U(1)^\prime_\eta$. For probability distributions P and Q defined on the same probability space, $\chi$, the Kullback-Leibler divergence from Q to P is defined to be
\begin{equation}
	D_{\rm KL} (P || Q) = \sum_{x \in \chi} P(x) log \left( \frac{P(x)}{Q(x)} \right)
\end{equation}
As seen from the bottom plane of Fig. \ref{fig:Zprime_pheno}, the Kullback-Leibler divergences are different in the vicinity of the $Z^\prime$ peak. In the bottom figure, we also introduce the  statistical uncertainty on the asymmetry, which is calculated through the formula 
\begin{equation}
	\delta A_{\rm FB} = \sqrt{ \frac{4}{ \mathcal{L}} \frac{\sigma_F \sigma_B}{ (\sigma_F + \sigma_B)^3}} = \sqrt{\frac{1 - A_{\rm FB}^2}{ \sigma \mathcal{L}}} =  \sqrt{\frac{1 - A_{\rm FB}^2}{ N }},
\end{equation}
where $\mathcal{L}$ is the integrated luminosity of the HL-LHC and $N$ the total number of events associated to it. The overarching message emerging from  the last plot of Fig. \ref{fig:Zprime_pheno} is that, alongside the total rate, $A_{\rm FB}$ is especially useful for all scenarios considered in order to separate these from the $U(1)_\psi$ and $U(1)^\prime_\eta$ model--dominated solutions. However, due to the limited number of events, some care should be given to optimising the binning, as  statistical uncertainties are rather large away from the $Z$-dominated solutions mass position, while they can well be controllable near it. In order to reduce the uncertainties in the asymmetry plot, for  the last plot of Fig. \ref{fig:Zprime_pheno}, we used variational bin size  (5.0-5.4, 5.4-5.6, 5.6-6.0 TeV), thus we  increased the number of events in each bin, which decreases the uncertainties. The conclusion from this plot is that, assuming that a $Z^\prime$ boson is experimentally discovered, then one can calculate the forward-backward asymmetry between 5.0 - 5.6 TeV, and distinguish our scenarios from $E_6$-motivated $U(1)^\prime_\eta$ and $U(1)^\prime_\psi$ models since the uncertainty bars do not cross each other. Unfortunately,  we cannot make this comment for $Z^\prime$ masses between 5.6 and 6.0 TeV, since there the uncertainty bars are still touching each other, in other words, the number of events is not enough in this region at 14 TeV and 3000 fb$^{-1}$.

\begin{figure}
	\centering
	\includegraphics[scale=0.28]{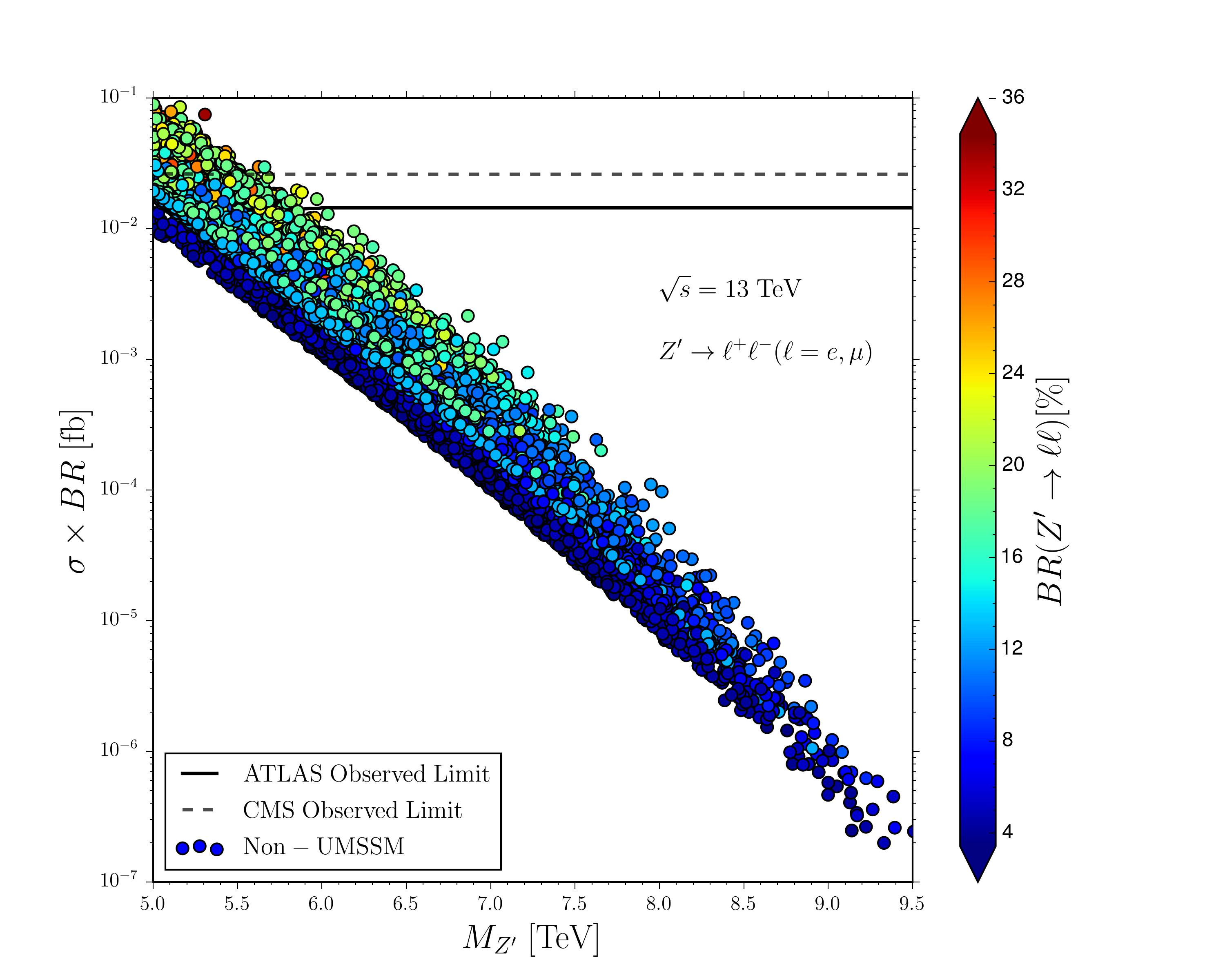}
	\includegraphics[scale=0.28]{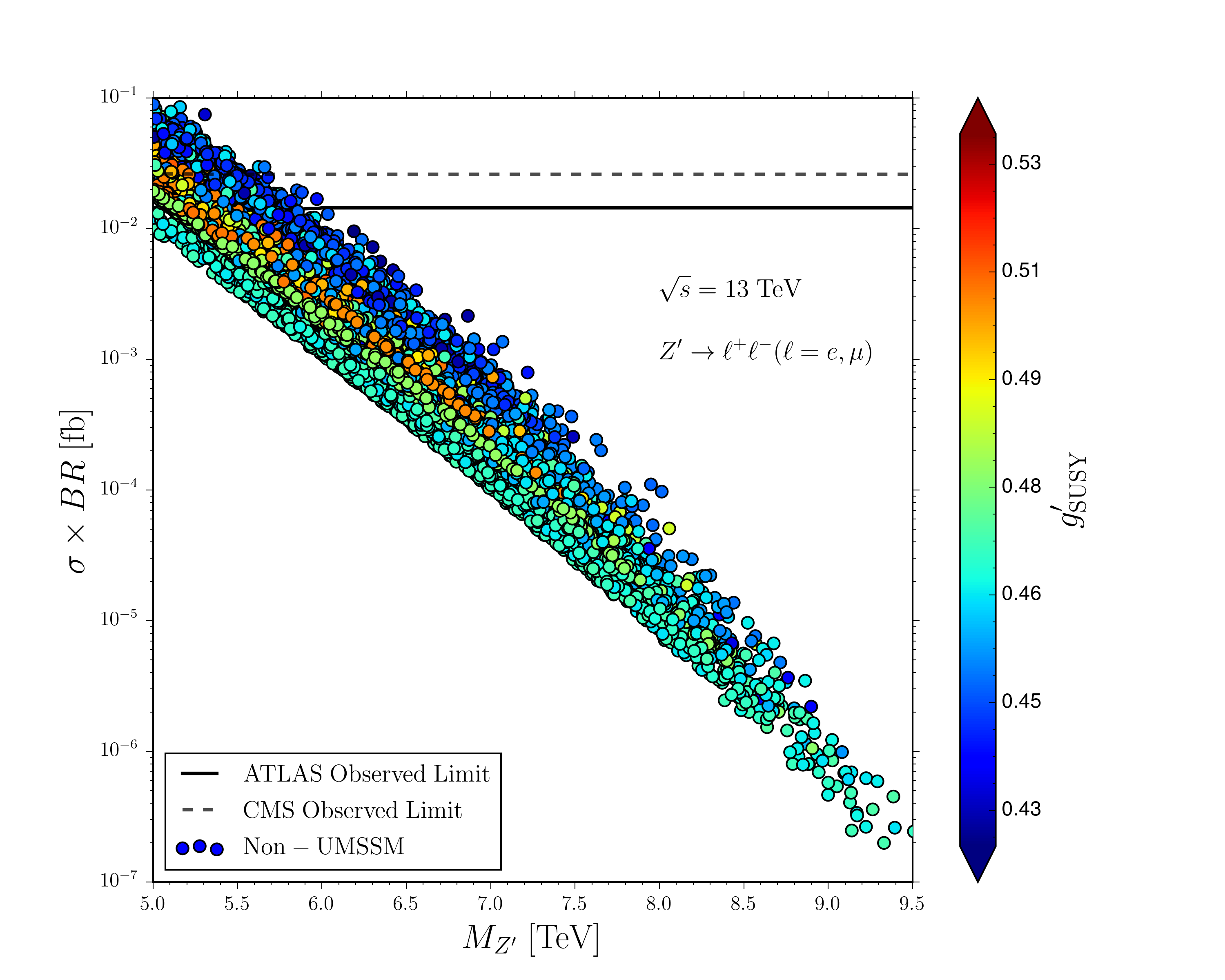} 
	\caption{$Z^\prime$ mass limits in the non-universal $U(1)^\prime$ model.}
	\label{fig:Zprime_limits}
\end{figure}

\begin{figure}
	\centering
	\includegraphics[scale=0.30]{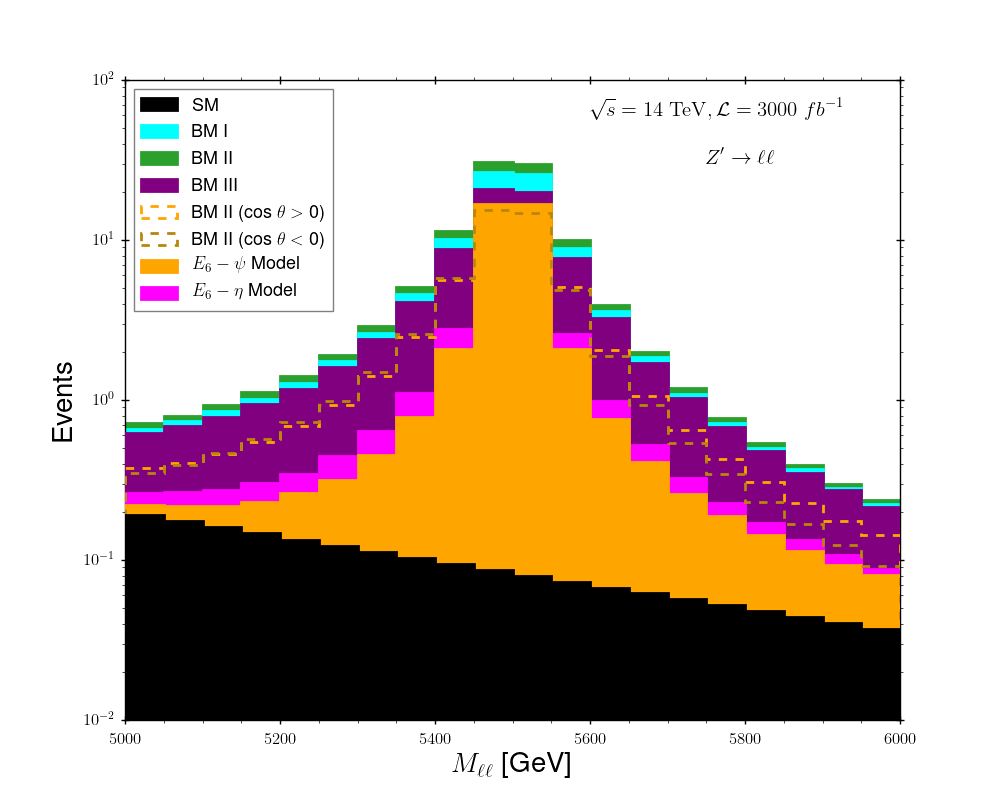}
	\includegraphics[scale=0.28]{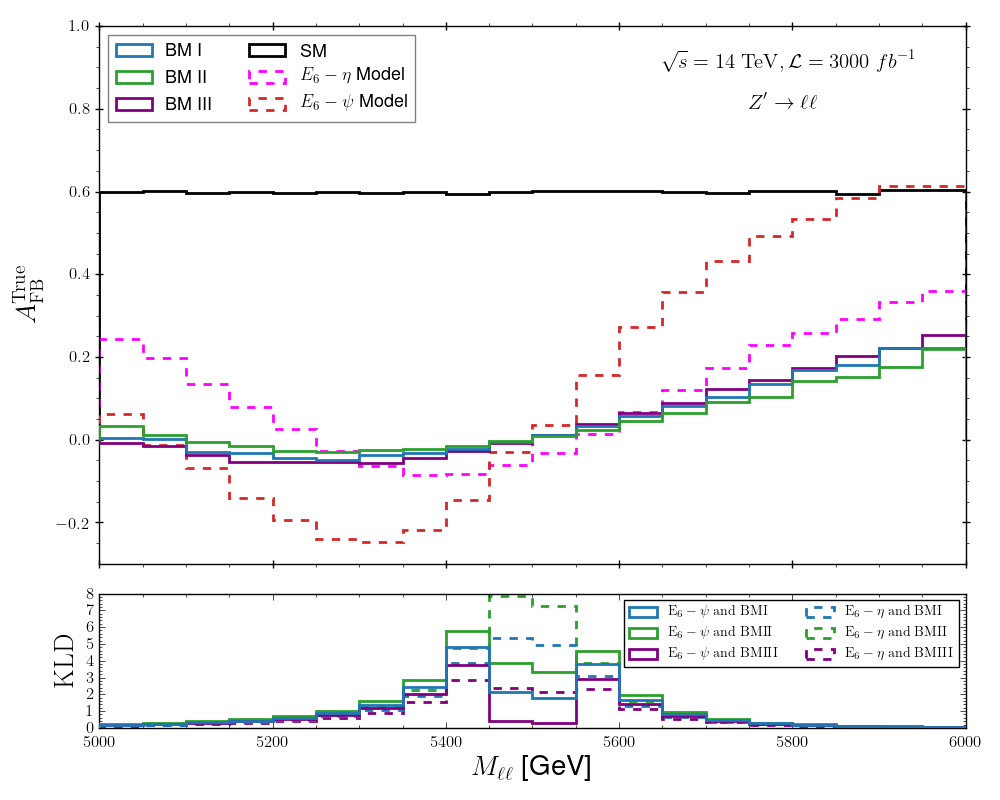} \\
	\includegraphics[scale=0.28]{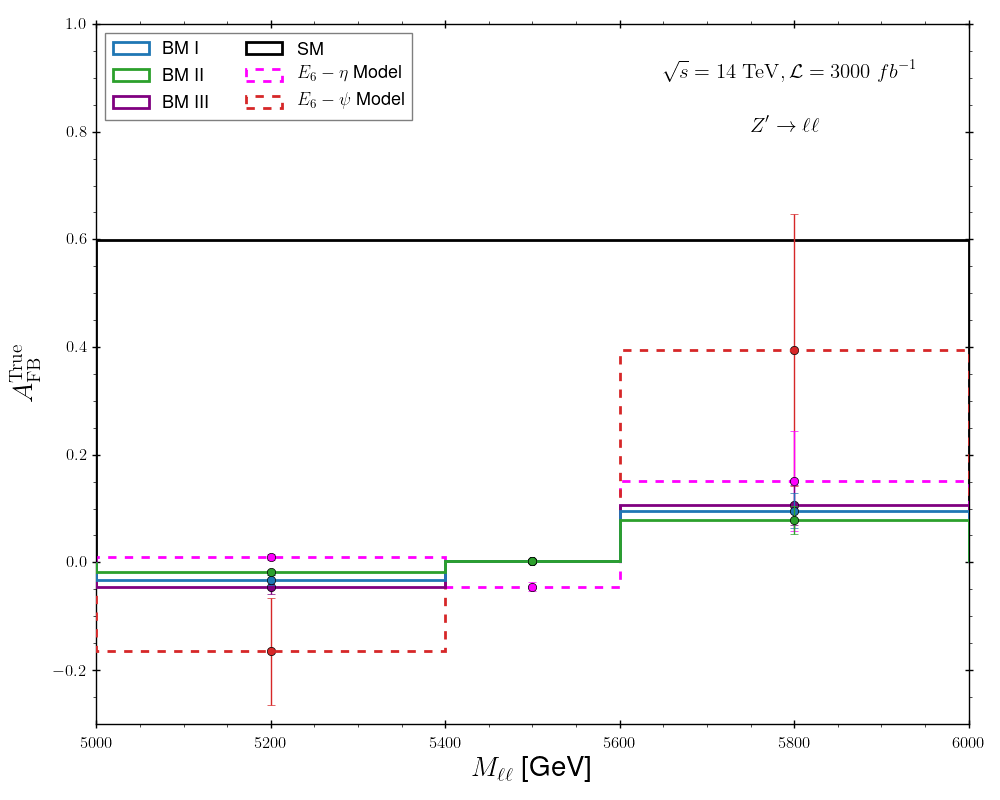}	
	\caption{$Z^\prime$ phenomenology at colliders. We depict signals from benchmarks \textbf{BM I} in  blue, \textbf{BM II} in green, and \textbf{BM III} in purple. Top left: Number of events in the NC DY channel versus the dilepton invariant mass within the SM, $E_6$ models of $\psi$ and $\eta$ type and \textbf{BM I}, \textbf{BM II} and \textbf{BM III}. For reference, \textbf{BM II} events are also given with cos $\theta > 0$ and  cos $\theta < 0$, where $\theta$ is the angle between the positively charged initial quark and negatively charged final state lepton.	Top right:  True $A_{\rm FB}$ versus the dilepton invariant mass within the SM, $E_6$ models of $\psi$ and $\eta$ type and  \textbf{BM I}, \textbf{BM II} and \textbf{BM III}. Results are given at the LHC with $\sqrt s=$ 14 TeV  and  $\mathcal{L} = $  3000 fb$^{-1}$ for bin widths equivalent to the expected mass resolution in the cross section distribution while at the bottom plane the results are optimized so as to distinguish between various scenarios.}
	\label{fig:Zprime_pheno}
\end{figure}

\section{Summary and Conclusion}
\label{sec:conclusion}
In this paper, we provide a solution to measurements of the anomalous magnetic moments for both the muon and the electron. We include supersymmetry to account for dark matter, and interpret the deviations as arising from beyond the MSSM scenarios.  As the ratio of the anomalous magnetic moments of electron and muon conflict assumptions about universal lepton gauge couplings (as the Yukawa couplings to the Higgs bosons are very small), we attempt a description within the simplest extension, that containing an additional Abelian gauge group, $U(1)^\prime$ with non-universal couplings. This model supplements the SM minimally by an additional neutral gauge boson $Z^\prime$, and by a singlet Higgs bosons which breaks $U(1)^\prime$ (and their fermionic partners).  The advantage of $U(1)^\prime$ models with non-universal couplings is that, in principle, anomaly cancellations can occur without the introduction of exotic states. This however means introduction of non-holomorphic terms in the Lagrangian, generating quark and/or lepton masses at loop level. Attempting to do so for the top quark is impossible, and for the bottom quark proves to be very difficult. We can circumvent the problem by introducing a minimum number of exotics, the so-called $D_x$ quarks and their partners, assumed to be heavy and decoupling from the spectrum. We then scan the model for a set of non-universal charges consistent with the anomaly conditions, non-zero charges for quarks and Higgs bosons, and correct ratios for the anomalous magnetic moments for electron and muon. These conditions are quite restrictive, and we are left with only 10 sets of  $U(1)^\prime$ charges. We further apply constraints from LHC involving $Z^\prime$ masses, chargino and neutralino masses, as well as dark matter constraints. The latter are applied to our analysis of dark matter, taken here to be the lightest neutralino, since sneutrino LSP scenarios are indistinguishable from those in MSSM plus a right-handed neutrino. We concentrate on the parameter space where anomalous magnetic moments are consistent with the measurements to 1$\sigma$, and choose three benchmarks: one where the LSP is bino-dominated, one where it is wino-dominated and one where it is higgsino-dominated. We  use these to then perform an analysis of the consequences of the model at the Run-3 at LHC. We compare some of our results with $E_6$ motivated $U(1)^\prime$ models with universal couplings. To do so, we calculate the number of events in the NC DY channel versus the dilepton invariant mass within the SM, $E_6$ models of $\psi$ and $\eta$ type and our three benchmark scenarios.  The branching ratios and production cross section of $Z^\prime$ are very similar in all models, and so our model cannot be distinguishable this way from other scenarios. In particular, the significance is low for both universal or non-universal scenarios. However, we find that the forward-backward asymmetry is very different for different models, and distinct in our model from $E_6$ motivated $U(1)^\prime_\eta$ and $U(1)^\prime_\psi$ models. Moreover, the asymmetry is the same for all three benchmarks thus this prediction for the model is very robust, and we show that the shape of the asymmetry is different, yielding hopes that it can be distinguished at the Run 3 at LHC.

\begin{acknowledgments}
 M.F. and \"{O}.\"{O}.  thank NSERC for partial financial support under grant number SAP105354. The work of YH is supported by The Scientific and Technological Research Council of Turkey (TUBITAK) in the framework of 2219-International Postdoctoral Research Fellowship Program. Parts  of  the  numerical  calculations  reported  in  this  paper  were  performed  using  High  Performance  Computing(HPC), managed by Calcul Quebec and Compute Canada, and the IRIDIS High Performance Computing Facility, and associated support services, at the University of Southampton. 
\end{acknowledgments}

\bibliography{MagMoment_combined}

\end{document}